\documentclass[10pt,A4paper]{article}
\usepackage{amssymb}
\usepackage{amsmath}
\usepackage{amsthm}
\usepackage{latexsym}
\usepackage[dvips]{epsfig}
\usepackage{mathrsfs}
\usepackage{eufrak}
\usepackage{bm}
\usepackage{tikz}
\usepackage{authblk}

\theoremstyle{plain}
\newtheorem{proposition}{Proposition}
\newtheorem{lemma}{Lemma}
\newtheorem{theorem}{Theorem}

\newtheorem*{main}{Theorem}

\def\bma{{\bm a}}
\def\bmb{{\bm b}}
\def\bmc{{\bm c}}
\def\bmd{{\bm d}}
\def\bme{{\bm e}}
\def\bmf{{\bm f}}
\def\bmg{{\bm g}}
\def\bmh{{\bm h}}
\def\bmi{{\bm i}}
\def\bmj{{\bm j}}
\def\bmk{{\bm k}}
\def\bml{{\bm l}}
\def\bmn{{\bm n}}
\def\bmm{{\bm m}}

\def\bms{{\bm s}}

\def\bmv{{\bm v}}

\def\bmx{{\bm x}}

\def\bmzero{{\bm 0}}
\def\bmone{{\bm 1}}
\def\bmtwo{{\bm 2}}
\def\bmthree{{\bm 3}}

\def\bmA{{\bm A}}
\def\bmB{{\bm B}}
\def\bmC{{\bm C}}
\def\bmD{{\bm D}}
\def\bmE{{\bm E}}
\def\bmF{{\bm F}}

\def\bmH{{\bm H}}
\def\bmK{{\bm K}}
\def\bmL{{\bm L}}

\def\bmN{{\bm N}}
\def\bmP{{\bm P}}
\def\bmQ{{\bm Q}}

\def\bmS{{\bm S}}
\def\bmT{{\bm T}}
\def\bmX{{\bm X}}

\def\mbfu{\mathbf{u}}

\def\bmalpha{{\bm \alpha}}

\def\bmgamma{{\bm \gamma}}

\def\bmepsilon{{\bm \epsilon}}

\def\bmxi{{\bm \xi}}

\def\bmomega{{\bm \omega}}

\def\bmsigma{{\bm \sigma}}

\def\bmUpsilon{{\bm \Upsilon}}

\def\bmpartial{{\bm \partial}}
\def\bmnabla{{\bm \nabla}}

\def\fraka{\mathfrak{p}}
\def\frakb{\mathfrak{q}}
\def\frakc{\mathfrak{r}}
\def\frakd{\mathfrak{s}}

\def\SigmaS{Z} 

\def\hS{l} 
\def\bmhS{\bml}
\def\KS{N}

\def\gsF{\mathcal{F}}

\newcounter{mnote}

\begin{document}

\title{\textbf{Spherically symmetric Anti-de Sitter-like Einstein-Yang-Mills spacetimes}}

\author[,1,2]{C. L\"ubbe \footnote{E-mail address:{\tt christian.luebbe@gmail.com}}}
\author[,3]{J. A. Valiente Kroon \footnote{E-mail address:{\tt j.a.valiente-kroon@qmul.ac.uk}}}
\affil[1]{Department of Mathematics, University College London, London, United Kingdom.}
\affil[2]{Graduate School of Mathematical Sciences, University of Tokyo, Tokyo, Japan.}
\affil[3]{School of Mathematical Sciences, Queen Mary, University of London, London, United Kingdom.}

\maketitle

\begin{abstract}
The conformal field equations are used to discuss the local existence
of spherically symmetric solutions to the Einstein-Yang-Mills system
which behave asymptotically like the anti-de Sitter spacetime. By
using a gauge based on conformally privileged curves we obtain a
formulation of the problem in terms of an initial boundary value
problem on which a general class of maximally dissipative boundary
conditions can be discussed. The relation between these boundary
conditions and the notion of mass on asymptotically anti-de Sitter
 spacetimes is analysed.
\end{abstract}



\section{Introduction}
\label{Introduction}

The purpose of this article is to give a first step towards the
construction, by means of conformal methods, of Einstein-Yang-Mills
spacetimes with anti-de Sitter-like boundary conditions. In this first analysis we
make the simplifying assumption of spherical symmetry, in order to
highlight various issues concerning the choice of boundary conditions
in the underlying initial boundary value problem.  The
construction of more general ---i.e. non-symmetric--- solutions to the
Einstein-Yang-Mills system with anti-de Sitter boundary conditions
will be treated elsewhere. Our interest in anti-de Sitter-like
spacetimes stems from the numerical results for the spherically symmetric Einstein-scalar
field system of 
\cite{BizRos11} which show evidence of turbulent instabilities in
arbitrarily small perturbations of the (vacuum) anti-de Sitter
spacetime for reflexive boundary conditions ---see also
e.g. \cite{Biz13,MalRos13} for a further discussion and further
references. As the aforementioned work shows, the assumption of
spherical symmetry offers a natural starting point to analyse anti de
Sitter-like spacetimes, not only from a numerical point of view but also from an
analytic perspective ---see e.g. \cite{HolSmu13} the stability of the
Schwarzschild-AdS spacetime for the spherically symmetric
Einstein-Klein-Gordon system.

\medskip
The construction of anti-de Sitter-like spacetimes is a much more
challenging problem than, say, the construction of de Sitter-like
spacetimes. These difficulties stem from the facts that anti-de Sitter-like
spacetimes are not globally hyperbolic and that
a systematic construction requires the formulation of an \emph{initial
boundary value problem} as well as the identification of suitable boundary
data. The first systematic construction of anti-de Sitter-like
spacetimes by means of an initial boundary value problem has been given
in \cite{Fri95}. This seminal work makes use the so-called \emph{conformal field
equations} together with a gauge based on the properties of
\emph{conformal geodesics} to show the \emph{local existence} of a
large class of anti-de Sitter-like spacetimes. Besides this analysis
for the vacuum case, one should also mention the work in
\cite{HolSmu12,HolWar13} in which the well-posedness of the
Einstein-scalar field with anti-de Sitter-like boundary
conditions has been analysed. 

\medskip
From our point of view, one of the key lessons of the analysis of
\cite{Fri95} is the derivation of the existence of a large class of
\emph{maximally dissipative} boundary conditions which ensure the
well-posedness of the underlying initial boundary value problem 
used for the construction of the anti-de Sitter-like spacetimes. In particular, these
maximally dissipative boundary conditions allow one to
prescribe the conformal class of the conformal boundary. In view of
this, it is natural to ponder how crucial the role of the reflexive
boundary conditions is in the formation of the instabilities observed in
\cite{BizRos11}. The set up considered in \cite{Fri95} is probably too
general to be used as a starting point for the analysis of the
stability/instability of the anti-de Sitter spacetime. From this
perspective the assumption of spherical symmetry seems a good way to
make inroads. As is well known, the trade-off for considering
spherically symmetric configurations in General Relativity is the need
of coupling the gravitational field to some matter model in order to
obtain non-trivial dynamics. If one intends to use conformal methods for the
analysis of this kind of problem, the choice of matter models narrows to
those possessing good conformal properties ---in particular Yang-Mills
fields or the conformally invariant scalar
field. The Maxwell field is of no use in spherical symmetry as it
leads to the Reissner-Nordstr\"om-anti-de Sitter spacetime and trivial
dynamics. In the present article we have opted to consider 
Yang-Mills fields as our matter model. From the point
of view of its conformal properties and its coupling to the conformal Einstein
field equations, the Yang-Mills field is better behaved than the conformally 
invariant scalar field. This outweighs the fact that for the the Yang-Mills field 
one has to consider more matter fields. The spherically symmetric Einstein-Yang-Mills
equations have been the subject of vast number of studies both
analytic ---see
e.g. \cite{BreForMai94,ForMan80,Kun91,KunMas90,Wit77}--- and numeric
---see e.g. \cite{BarMcK88,Biz90,ChoChmBiz96,RinMon13}. It is
imporatnt to point out that in these
studies the gauge freedom available in the specification of the gauge
potential has been systematically used to eliminate one of the
components of this field, and thus, to obtain a simpler system of
equations to be solved. In this article the gauge freedom is used in a
different manner to obtain a symmetric hyperbolic evolution system for the
components of the gauge potential with nice properties at the
conformal boundary. This different strategy in the use of the gauge
freedom makes difficult a direct general comparison between the present
analysis and other set ups ---although a case by case comparison can
be certainly obtained if required. Another point that should be
stressed is that while most of the analysis of spherically symmetric
Einstein-Yang-Mills spacetimes has been carried out for the in which
the gauge group is $SU(2)$, \emph{our analysis is completely general and
makes no assumptions on the gauge group}. 

\medskip 
In the present article we pursue a generalisation of the analysis in
\cite{Fri95} by combining it with the discussion of the conformal 
Einstein-Yang-Mills equations given in \cite{Fri91} and that of
\cite{LueVal12} on the so-called \emph{extended conformal Einstein 
field equations} for Einstein-Maxwell. Our strategy is to express the 
extended conformal Einstein field equations in terms of a gauge based on
the properties of certain conformally privileged curves, the so-called
\emph{conformal curves}, and in turn formulate a boundary value problem 
including the prescription of certain information on the conformal boundary
$\mathscr{I}$. The solutions obtained from this initial boundary value
problem are, in principle, global in space but local in time. It is well known that 
the assumption of spherical symmetry leads to a reduction in the number of 
evolution equations. In addition, this setting exhibits further structural properties 
which in our opinion deserve a separate treatment. 
One of the key objectives of the analysis in this article is
to identify the boundary data to be prescribed at the conformal
boundary in order to ensure the (local) existence of an anti-de
Sitter-like Einstein-Yang-Mills spacetime.  Our analysis shows that
the construction of a Einstein-Yang-Mills spacetime by means of an
initial boundary value problem allows to specify a certain combination
of the components of the gauge field. More precisely, if $A^\fraka_-$
and $A^\fraka_+$ denote the null components of the gauge fields with
respect to a frame adapted to the conformal boundary, then
\begin{equation}
A^\fraka_-  = c^\fraka A^\fraka_+ + q^\fraka, \qquad -1 \le c^\fraka
\leq 1 \qquad \textmd{for each } \fraka
\label{Introduction:BoundaryConditions}
\end{equation}
with $q^\fraka$ a collection of smooth functions on the conformal
boundary, 
constitutes a suitable class of \emph{maximally dissipative boundary
  conditions} ensuring the well-posedness of the underlying initial
boundary value problem. In the above expression $\fraka$ denotes a suitable \emph{gauge index}  and no
summation is understood on the repeated indices. Our main result can
be expressed as follows:

\begin{main}
Suppose one is given smooth spherically symmetric anti-de Sitter-like
initial data for the Einstein-Yang-Mills equations on a 3-dimensional
manifold $\mathcal{S}$ with boundary $\partial\mathcal{S}$. Suppose further
that the gauge potentials for the Yang-Mills field satisfy the boundary condition
\eqref{Introduction:BoundaryConditions} on a cylinder
$[0,T]\times \partial \mathcal{S}$ for some $T>0$, and that their initial
data on $\mathcal{S}$ satisfies certain compatibility conditions at the corner
$\{0\}\times \partial\mathcal{S}$. Then there exists a local-in-time
solution to the Einstein-Yang-Mills equations with an anti-de
Sitter-like cosmological constant which possesses a conformal completion
such that on $\{0\}\times\mathcal{S}$ it implies the given initial
data. Moreover, this solution to the Einstein field equations admits a
conformal completion on $\mathcal{M}_T\equiv [0,T]\times\mathcal{S}$
such that $\mathscr{I}^+\equiv [0,T]\times \partial \mathcal{S} $
corresponds to conformal boundary and the boundary conditions
\eqref{Introduction:BoundaryConditions} are satisfied on
$\mathscr{I}^+$.
\end{main}

A detailed discussion of the assumptions made in the above result will
be given in the main text ---see Section \ref{Section:Main result},
Theorem \ref{Theorem:MainTheoremPreciseVersion}. In particular, in
Section \ref{Section:Evolution equs SS} the meaning of spherical
symmetry in the present context will be made precise. The
conditions at the corner $\{0\}\times \partial
\mathcal{S}$ (corner conditions) are a hierarchy of compatibility
conditions between the boundary data and the initial data ensuring the
smoothness of the solution to the initial boundary value
problem. Their precise form is discussed in Section
\ref{Section:CornerConditions}. It should also be pointed out that in
order to close the problem it is necessary to specify an, in
principle, arbitrary gauge source function $F^\fraka(x)$ on
$\mathcal{M}_T$ which fixes the value of the divergence of
the Yang-Mills gauge potential 1-form. The specification of this gauge
source function must be made in a manner which is consistent with
boundary conditions. This consistency requirement is part of the
corner conditions.

\begin{figure}[t]
\begin{center}
\includegraphics[scale=1.2]{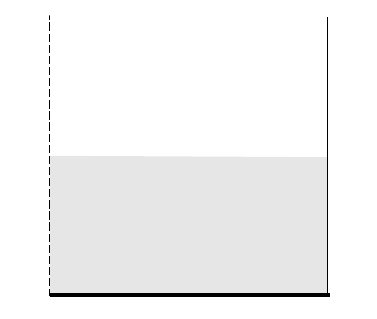}
\put(-65,0){$\mathcal{S}$}
\put(-15,70){$\mathscr{I}$}
\end{center}
\caption{Penrose diagram of the Einstein-Yang-Mills spacetime with
anti-de Sitter-like boundary conditions obtained in the present
article. The sets $\mathcal{S}$ and $\mathscr{I}$ denote, respectively
an initial hypersurface and the conformal boundary. The spacetime
obtained as a result of our main theorem (indicated by the grey shaded
region) is global in space but local in time.}
\label{Figure:MainResult}
\end{figure}

\medskip
 A schematic representation of the result is
given in the Penrose diagram of Figure \ref{Figure:MainResult}. The
proof of the above result requires a careful specification of the
gauge used to extract an evolution system out of the conformal field
equations. In order to avoid a \emph{free boundary problem}, we resort
to a conformal gauge based on the properties of a class of curves with
nice conformal properties, the so-called conformal curves ---see
e.g. \cite{LueVal12}. A key feature of this gauge is that the
conformal factor is known \emph{a priori} since it can be expressed 
explicitly in terms of initial data. Accordingly,
the location of the conformal boundary is explicitly known as well. This
feature considerably simplifies the problem and reduces complexity of the PDE theory
required to show well-posedness.   

\medskip
Concerning the above result, it should be pointed out that the
boundary conditions \eqref{Introduction:BoundaryConditions} contain,
as a particular case, the \emph{reflective boundary conditions}
\[
A^\fraka_- = A^\fraka_+.
\]
A further interesting property of the setting to be considered
concerns the behaviour of the mass. As it will be discussed in more
detail in the main text, the spherical symmetry of the setting forces
the geometry of the metric intrinsic to the conformal boundary to be
conformally flat. That is, the boundary metric is conformally related
to the 1+2-dimensional analogues of the Minkowski metric and the
metric of the 1+2-dimensional Einstein cylinder universe. For anti-de
Sitter-like spacetimes with this property there exists a well defined
notion of mass ---see \cite{AshMag84}. A direct computation shows that
for the class of spherically symmetric Einstein-Yang-Mills spacetimes
constructed in this article the mass is, in general, not constant
along the conformal boundary.

\subsection*{Outline of the article}
In Section \ref{Section:ConformalEinsteinYangMillsEquations} we start
by discussing background material concerning the Einstein-Yang-Mills
field equations in a conformal setting.  Section
\ref{Section:GeneralProperties} provides an overview of general
properties of spherically symmetric anti-de Sitter-like Yang-Mills
spacetimes. In particular, this section provides a discussion of the
solutions to the constraints implied by the conformal field equations
at the conformal boundary of the spacetime and of the behaviour of the
mass of the spacetime.  Section \ref{Section:Spacetime gauge
considerations} introduces the concept of conformal curves and studies
their properties at $\mathscr{I}$. In particular, Lemma
\ref{Lemma:ConformalCurvesConformalBoundary} proves the existence of
conformal curves that remain tangent to the conformal boundary
$\mathscr{I}$ for all times.  A congruence of conformal curves is used
in Section \ref{Section:IBVP} to construct a boundary adapted
coordinate system which, in turn, is employed to formulate the initial boundary
value problem. Moreover, the congruence is used to suitably fix the
conformal gauge freedom so that a hyperbolic reduction
of the conformal Einstein-Yang-Mills field equations can be carried out. The equations
are analysed in the setting of spherical symmetry in Section
\ref{Section:Evolution equs SS} and in Section \ref{Section:boundary
conditions} the boundary conditions for the gauge potential $A^\fraka$
are discussed.  The main result on the existence of spherically
symmetric Einstein-Yang-Mills spacetimes is proven in Section
\ref{Section:Main result}, which also presents the technical version
of our main theorem. Some concluding remarks are given in Section
\ref{Section:Discussion}.

\subsection*{Notation and Conventions}

Our signature convention for spacetime (Lorentzian) metrics is
$(+---)$.  As a consequence of this signature convention, 
the cosmological constant of anti-de Sitter like spacetimes is
positive. 

In what follows ${}_a,\, {}_b, \, {}_c, \ldots$ denote spacetime
tensorial indices while ${}_\bma,\, {}_\bmb, \, {}_\bmc, \ldots$
correspond to spacetime frame indices taking the values
$0,\ldots,3$. For the geometry of a 3-dimensional submanifold the
tensorial indices will be denoted by ${}_i, \,{}_j,\, {}_k, \ldots$
and frame indices by ${}_\bmi, \, {}_\bmj, \, {}_\bmk$. The frame
indices take values $\mathbf{0}, \mathbf{1}, \mathbf{2} $ for timelike
submanifolds and values $\mathbf{1}, \mathbf{2}, \mathbf{3} $ for
spacelike submanifolds.  Part of the analysis will require the use of
spinors. In this respect we make use of the general conventions of
Penrose \& Rindler \cite{PenRin84}. In particular, ${}_A, \,{}_B,
\,{}_C, \dots$ denote abstract spinorial indices, while ${}_\bmA,
\,{}_\bmB, \,{}_\bmC, \dots$ indicate frame spinorial indices with
respect to some specified spin dyad $\{ \delta_\bmA \}$.

Wherever we consider it preferable for readability, we will suppress
tensorial indices and write the corresponding tensor in a bold
font. Examples for this are the metric $\bmg$, the frame basis $\bme_\bma$
and dual cobasis $\bmomega^\bma$.

Various connections will be used throughout. The connection
$\tilde{\bmnabla}$ will always denote the Levi-Civita connection of a
Lorentzian metric $\tilde{\bmg}$ satisfying the Einstein-Yang-Mills field
equations ---hence, we call it the \emph{physical connection}. 
$\bmnabla$ will denote the Levi-Civita connection of a conformally related 
\emph{unphysical metric} $\bmg$, while $\hat{\bmnabla}$ will 
denote a Weyl connection of the conformal class $[\tilde{\bmg}] $.

\section{The conformal Einstein-Yang-Mills equations}
\label{Section:ConformalEinsteinYangMillsEquations}

In what follows let $(\tilde{\mathcal{M}},\tilde{\bmg},\mathfrak{G},\bmF^\fraka,\bmA^\fraka)$ denote a
spacetime satisfying the Einstein-Yang-Mills field equations. These field
equations provide differential conditions for the spacetime metric
$\tilde{g}_{ab}$ and a set of antisymmetric \emph{gauge fields}
$F^\fraka{}_{ab}$ and \emph{gauge potential 1-forms}
$A^\fraka{}_a$ where the indices
$\fraka,\;\frakb,\cdots$ take values in a Lie algebra $\mathfrak{g}$
of a Lie group $\mathfrak{G}$. Explicitly, the field equations are given by 
\begin{subequations}
\begin{eqnarray}
&& \tilde{R}_{ab} - \tfrac{1}{2} \tilde{R} \tilde{g}_{ab} + \lambda
\tilde{g}_{ab} = \tilde{T}_{ab}, \label{TensorialEinstein}\\
&& \tilde{\nabla}_a A^\fraka{}_b - \tilde{\nabla}_b A^\fraka{}_a +
C^\fraka{}_{\frakb\frakc} A^\frakb{}_a A^\frakc{}_b -F^\fraka{}_{ab} =0, \label{TensorialYM1}\\
&& \tilde{\nabla}^a F^\fraka{}_{ab} + C^\fraka{}_{\frakb\frakc}
A^{\frakb a} F^\frakc_{ab}=0, \label{TensorialYM2}
\end{eqnarray}
\end{subequations}
where $\lambda$ is the cosmological constant and the energy-momentum of the Yang-Mills
field is given by
\[
\tilde{T}_{ab} = \frac{1}{4}
\delta_{\fraka\frakb} F^\fraka{}_{cd} F^{\frakb cd} g_{ab}
-\delta_{\fraka\frakb}F^\fraka{}_{ac}F^\frakb{}_b{}^c.
\]
In equation \eqref{TensorialEinstein} $\tilde{\nabla}_a$,
$\tilde{R}_{ab}$ and $\tilde{R}$ denote, respectively,  the
Levi-Civita connection, Ricci tensor and Ricci scalar of the metric
$\tilde{g}_{ab}$ while in equations
\eqref{TensorialYM1}-\eqref{TensorialYM2} $C^\fraka{}_{\frakb\frakc}=C^\fraka{}_{[\frakb\frakc]}$ denote
the structure constants of the Lie algebra $\mathfrak{g}$ and 
$\delta_{\fraka\frakb}$ is the Kronecker delta. Our
analysis will make no restriction on the nature of this Lie
algebra. It is important to notice that there is a certain gauge
freedom in the specification of the Yang-Mills gauge potential
1-form. More precisely, $A^\fraka{}_a$ is determined up to the
gradient of a scalar function. In Section \ref{Subsection:HR_EYMEqns}, this gauge freedom
will be used to make the divergence of $A^\fraka{}_a$ equal to an
arbitrary $\mathfrak{g}$-valued function $\gsF^\fraka(x)$ on the spacetime
manifold. 

The gauge fields $F{}^\fraka_{ab}$ satisfy the \emph{Bianchi identity}
\[
\tilde{\nabla}_{[a} F{}^\fraka_{bc]} + C^\fraka{}_{\frakb\frakc}
A^\frakb{}_{[a} F^\frakc{}_{bc]}=0.
\]
By introducing the \emph{Hodge dual} $F^{\fraka*}_{ab}$, one can
rewrite the last equation so that it resembles equation
\eqref{TensorialYM2}:
\begin{equation}
\tilde{\nabla}^a F^{\fraka*}{}_{ab} + C^\fraka{}_{\frakb\frakc}
A^{\frakb a} F^{\frakc*}{}_{ab}=0.
\label{TensorialYM3}
\end{equation}

\medskip
The Yang-Mills equations \eqref{TensorialYM1}-\eqref{TensorialYM2} are
conformally invariant. More precisely, suppose we are given the unphysical metric
$g_{ab}=\Xi^2 \tilde{g}_{ab}$ for some conformal factor $\Xi$ and 
let $\nabla_a$ denote the Levi-Civita connection of the metric $g_{ab}$.
If the fields $F^\fraka{}_{ab}$ and $A^\fraka{}_a$ are solutions to equations
\eqref{TensorialYM1}-\eqref{TensorialYM2} for a given
$\tilde{g}_{ab}$, then they also solve the equations
\begin{eqnarray*}
&& \nabla_a A^\fraka{}_b - \nabla_b A^\fraka{}_a +
C^\fraka{}_{\frakb\frakc} A^\frakb{}_a A^\frakc{}_b -F^\fraka{}_{ab} =0, \\
&& \nabla^a F^\fraka{}_{ab} + C^\fraka{}_{\frakb\frakc}
A^{\frakb a} F^\frakc_{ab}=0, 
\end{eqnarray*}
 For future use it is convenient to define the
\emph{unphysical energy-momentum tensor} $T_{ab}$ as
\[
T_{ab} \equiv \Xi^{-2} \tilde{T}_{ab}.
\]

Note that the Einstein equations are not conformally invariant and hence 
$T_{ab}$ and $R_{ab}$ are not related by an analogue version of
\eqref{TensorialEinstein} for the unphysical metric $\bmg$. This
observation naturally leads to consider the conformal Einstein field
equations.

\subsection{The conformal Einstein field equations with matter}
In the present analysis we will make use of a formulation of the
conformal field equations expressed in terms of Weyl connections
---i.e. a torsion-free connection, not necessarily a Levi-Civita connection,
which preserves the conformal structure. This general version of the
conformal equations is called the \emph{extended conformal Einstein field equations}. These
equations were originally introduced in \cite{Fri91} for the vacuum
case and in \cite{LueVal12} for the case of matter with trace-free
energy-momentum tensor. 

\medskip
For completeness, we briefly review the general setting of the
extended conformal field equations for a trace-free
energy-momentum tensor. As in the previous section, given
a conformal factor we define the \emph{unphysical metric} $\bmg$ by
the relation
\[
\bmg = \Xi^2 \tilde{\bmg}.
\]
 Let $\{\bme_\bma\}$,
$\bma=0,\ldots,3$ denote a frame field which is $\bmg$-orthogonal so
that $\bmg(\bme_\bma,\bme_\bmb) =\eta_{\bma\bmb}$, and let $\{
\bmomega^\bmb \}$ denote its dual cobasis ---i.e. $\langle
\bmomega^\bmb, \bme_\bma \rangle=\delta_\bma{}^\bmb$. 
The connection coefficients $\Gamma_\bma{}^\bmc{}_\bmb =\langle
\bmomega^\bmc, \nabla_\bma \bme_\bmb\rangle$ of the 
Levi-Civita connection ${\bm \nabla}$ of $\bmg$ satisfy the usual metric
compatibility condition $\Gamma_\bma{}^\bmb{}_\bmb=0$. Given a smooth 
1-form $\bmf$ one can
define a Weyl connection $\hat{\bmnabla}$ through the relation 
\begin{equation}
\label{Connection change}
\hat{\Gamma}_\bma{}^\bmc{}_\bmb =  \Gamma_\bma{}^\bmc{}_\bmb + \delta _\bma{}^\bmc f_\bmb
+ \delta _\bmb{}^\bmc f_\bma -\eta_{\bma\bmb}\eta^{\bmc\bmd}f_\bmd,
\end{equation}
where $\hat{\Gamma}_\bma{}^\bmc{}_\bmb =\langle
\bmomega^\bmc, \hat{\nabla}_\bma \bme_\bmb\rangle$. In particular, one has that $f_\bma =
\tfrac{1}{4}\hat{\Gamma}_\bma{}^\bmb{}_\bmb$. Hence $\Gamma_\bma{}^\bmc{}_\bmb$ can be fully expressed in terms of $\hat{\Gamma}_\bma{}^\bmc{}_\bmb$ using \eqref{Connection change}.

\medskip
It is convenient to
distinguish between the expression for the components of the Riemann
tensor of the connection $\hat{\bm \nabla}$ in terms of the connection
coefficients $\hat{\Gamma}_\bma{}^\bmc{}_\bmb$ (the \emph{geometric
curvature} $\hat{P}^\bmc{}_{\bmd\bma\bmb}$) and the expression of the
Riemann tensor in terms of the Schouten and Weyl tensors
(the \emph{algebraic curvature}
$\hat{\rho}^\bmc{}_{\bmd\bma\bmb}$). Explicitly, one has that
\begin{eqnarray*}
&& \hat{P}^\bmc{}_{\bmd\bma\bmb} \equiv \bme_\bma(\hat{\Gamma}_\bmb{}^\bmc{}_\bmd) -
\bme_\bmb(\hat{\Gamma}_\bma{}^\bmc{}_\bmd) \\
&& \hspace{2cm} + \hat{\Gamma}_\bmf{}^\bmc{}_\bmd(
\hat{\Gamma}_\bmb{}^\bmf{}_\bma - \hat{\Gamma}_\bma{}^\bmf{}_\bmb ) +
\hat{\Gamma}_\bmb{}^\bmf{}_\bmd \hat{\Gamma}_\bma{}^\bmc{}_\bmf  -
\hat{\Gamma}_\bma{}^\bmf{}_\bmd \hat{\Gamma}_\bmb{}^\bmc{}_\bmf, \\
&& \hat{\rho}^\bmc{}_{\bmd\bma\bmb} \equiv \Xi d^\bmc{}_{\bmd\bma\bmb}
+ 2 (\delta^\bmc{}_{[\bma} \hat{L}_{\bmb]\bmd} - \delta^\bmc{}_{\bmd} \hat{L}_{[\bma\bmb]} - g_{\bmd[\bma} \hat{L}_{\bmb]}{}^\bmc),
\end{eqnarray*}
where $d^\bmc{}_{\bmd\bma\bmb}\equiv \Xi^{-1}
C^\bmc{}_{\bmd\bma\bmb}$ denotes the components of the rescaled
Weyl tensor with respect to the frame $\{ \bme_\bma \}$ and
$\hat{L}_{\bma\bmb}$ those of the Schouten tensor of the connection
$\hat{\bmnabla}$.

\medskip
In order to write down the conformal field equations, it is convenient
to define the \emph{geometric zero-quantities}
\begin{eqnarray*}
&& \hat{\Sigma}_\bma{}^\bmc{}_\bmb \bme_\bmc \equiv  [\bme_\bma,\bme_\bmb] 
-(\hat{\Gamma}_\bma{}^\bmc{}_\bmb-\hat{\Gamma}_\bmb{}^\bmc{}_\bma)
\bme_\bmc, \\
&& \hat{\Xi}^\bmc{}_{\bmd\bma\bmb} \equiv \hat{P}^\bmc{}_{\bmd\bma\bmb} -
{\rho}^\bmc{}_{\bmd\bma\bmb}, \\
&& \hat{\Delta}_{\bmc\bmd\bmb} \equiv \hat{\nabla}_\bmc \hat{L}_{\bmd\bmb} - \hat{\nabla}_\bmd \hat{L}_{\bmc\bmb} - d_\bma d^\bma{}_{\bmb\bmc\bmd} - \Xi
T_{\bmc\bmd\bmb}, \\
&&  \Lambda_{\bmb\bmc\bmd}\equiv  \nabla_\bma
d^\bma{}_{\bmb\bmc\bmd} -
T_{\bmc\bmd\bmb},
\end{eqnarray*}
and the \emph{matter zero-quantities}
\begin{eqnarray*}
&& M^\fraka{}_{\bma\bmb}\equiv \nabla_\bma A^\fraka{}_\bmb - \nabla_\bmb A^\fraka{}_\bma +
C^\fraka{}_{\frakb\frakc} A^\frakb{}_\bma A^\frakc{}_\bmb -F^\fraka{}_{\bma\bmb},\\
&& M^\fraka{}_\bmb \equiv \nabla^\bma F^\fraka{}_{\bma\bmb} + C^\fraka{}_{\frakb\frakc}
A^{\frakb \bma} F^\frakc_{\bma\bmb}, \\
&& M^{\fraka*}{}_\bmb \equiv \nabla^\bma F^{\fraka*}{}_{\bma\bmb} + C^\fraka{}_{\frakb\frakc}
A^{\frakb a} F^{\frakc*}{}_{\bma\bmb},
\end{eqnarray*}
where $T_{\bmc\bmd\bmb} \equiv \Xi^{-1} \tilde{\nabla}_{[\bmc} \tilde{T}_{\bmd]\bmb}$ denotes the \emph{rescaled Cotton-York tensor}. Expressed it in terms of the unphysical Levi-Civita connection and the unphysical energy-momentum tensor $T_{ab}$ one has that
\begin{equation}
T_{\bma\bmb\bmc} = \Xi \nabla_{[\bma} T_{\bmb]\bmc} +3 \nabla_{[\bma} \Xi T_{\bmb]\bmc} - g_{\bmc[\bma} \nabla^\bme \Xi T_{\bmb]\bme}. 
\label{RescaledCottonYork}
\end{equation}
The fields $f_\bma$, $d_\bma$ and $\Xi$ are related to each other by
the constraint
\[
d_\bma = f_\bma + \nabla_\bma \Xi.
\]
This last expression can be used in formula \eqref{RescaledCottonYork}
to eliminate the gradient of the conformal factor.

\medskip
\noindent 
\textbf{Remark.} The geometric zero-quantity $\hat{\Sigma}_\bma{}^\bmc{}_\bmb$ can be viewed as the torsion of the connection $\hat{\bm\nabla}$.
In the last geometric zero-quantity and in the matter zero-quantities we have used the unphysical Levi-Civita connection $\bmnabla$. This has been done to ease readability of these equations. As mentioned before, the connection coefficients of $\bmnabla$ can be expressed entirely in terms of $\hat{\Gamma}_\bma{}^\bmc{}_\bmb$ using \eqref{Connection change}. The equations in terms of $\hat{\bmnabla}$ can be found in \cite{LueVal12}.

\medskip
In terms of the above zero-quantities the \emph{extended conformal
Einstein-Maxwell field equations} are given by the conditions
\begin{subequations}
\begin{eqnarray}
&\hat{\Sigma}_\bma{}^\bmc{}_\bmb\bme_\bmc=0, \qquad \hat{\Xi}^\bmc{}_{\bmd\bma\bmb}=0,
\qquad \hat{\Delta}_{\bmc\bmd\bmb} =0, \qquad \Lambda_{\bmb\bmc\bmd} =0 & \label{XCFEFrame1}\\
& M^\fraka{}_{\bma\bmb}=0, \qquad M^\fraka{}_\bmb=0, \qquad
M^{\fraka*}{}_\bmb=0. \label{XCFEFrame2}
\end{eqnarray}
\end{subequations}

The above conformal equations can be read as yielding differential
conditions, respectively, for the frame components $e_\bma{}^a$, the
spin coefficients $\hat{\Gamma}_\bma{}^\bmc{}_\bmb$ (including the the
components $f_\bma$ of the 1-form $\bmf$), the components of the
Schouten tensor $\hat{L}_{\bma\bmb}$, the components of the rescaled
Weyl tensor $d^\bma{}_{\bmb\bmc\bmd}$, and the collection of matter fields
$F^\fraka{}_{ab}$ and $A^\fraka{}_a$.

\medskip
\noindent
\textbf{Remark.} The conformal equations
\eqref{XCFEFrame1}-\eqref{XCFEFrame2} have to be supplemented with
gauge conditions or equations which determine the conformal factor
$\Xi$ and the 1-form $\bmd$. This will be discussed in Section \ref{Section:GaugeConditions}.

\subsubsection{Spinorial formulation of the equations}
 The spinorial counterparts of the fields
\[
\bme_\bma, \quad \hat{\Gamma}_\bma{}^\bmb{}_\bmc, \quad f_\bma,\quad
\hat{L}_{\bma\bmb}, \quad d^\bma{}_{\bmb\bmc\bmd}, \quad d_\bma,\quad  T_{\bma\bmb\bmc}, 
\]
are given, respectively, by the spinor fields
\begin{equation}
\bme_{\bmA\bmA'}, \quad \hat{\Gamma}_{\bmA\bmA'\bmB\bmC},\quad f_{\bmA\bmA'}, \quad
\hat{L}_{\bmA\bmA'\bmB\bmB'}, \quad \phi_{\bmA\bmB\bmC\bmD}, \quad d_{\bmA\bmA'}, \quad T_{\bmA\bmB\bmC\bmC'},
\label{SpinorialUnknowns}
\end{equation}
with 
\[
 \phi_{\bmA\bmB\bmC\bmD}=
\phi_{(\bmA\bmB\bmC\bmD)}, \qquad T_{\bmA\bmB\bmC\bmC'} = T_{(\bmA\bmB)\bmC\bmC'}, 
\]
The spinorial counterpart of the zero-quantities encoding the conformal Einstein-Yang-Mills field equations is given by
\begin{subequations}
\begin{eqnarray}
&& \hat{\Sigma}_{\bmA\bmA'}{}^{\bmC\bmC'}{}_{\bmB\bmB'} \bme_{\bmC\bmC'}\equiv
[\bme_{\bmA\bmA'},\bme_{\bmB\bmB'}]
-(\hat{\Gamma}_{\bmA\bmA'}{}^{\bmC\bmC'}{}_{\bmB\bmB'}-\hat{\Gamma}_{\bmB\bmB'}{}^{\bmC\bmC'}{}_{\bmA\bmA'})\bme_{\bmC\bmC'},
\label{XCFESpinorZQ1}\\
&& \hat{\Xi}^{\bmC}{}_{\bmD\bmA\bmA'\bmB\bmB'} \equiv \hat{P}^{\bmC}{}_{\bmD\bmA\bmA'\bmB\bmB'} -
\hat{\rho}^{\bmC}{}_{\bmD\bmA\bmA'\bmB\bmB'}, \label{XCFESpinorZQ2}\\
&& \hat{\Delta}_{\bmC\bmC'\bmD\bmD'\bmB\bmB'} \equiv \hat{\nabla}_{\bmC\bmC'}
\hat{L}_{\bmD\bmD'\bmB\bmB'} - \hat{\nabla}_{\bmD\bmD'} \hat{L}_{\bmC\bmC'\bmB\bmB'}\nonumber  \\
&&\hspace{3cm} -
d^{\bmA\bmA'}(\phi_{\bmA\bmB\bmC\bmD}\epsilon_{\bmA'\bmB'}\epsilon_{\bmC'\bmD'} - \bar{\phi}_{\bmA'\bmB'\bmC'\bmD'} \epsilon_{\bmA\bmB} \epsilon_{\bmC\bmD})\nonumber \\
&& \hspace{3cm} - \Xi( T_{\bmC\bmD\bmB\bmB'}\epsilon_{\bmC'\bmD'} + \bar{T}_{\bmC'\bmD'\bmB'\bmB}\epsilon_{\bmC\bmD}), \label{XCFESpinorZQ3}\\
&&  \Lambda_{\bmA'\bmA\bmB\bmC}\equiv \nabla^\bmQ{}_{\bmA'} \phi_{\bmA\bmB\bmC\bmQ} - T_{\bmB\bmC\bmA\bmA'}
 \label{XCFESpinorZQ4},
\end{eqnarray}
\end{subequations}
and the reduced geometric and algebraic curvature zero-quantities are given by
\begin{eqnarray*}
&&  \hat{P}^{\bmC}{}_{\bmD\bmA\bmA'\bmB\bmB'} \equiv
\bme_{\bmA\bmA'}(\hat{\Gamma}_{\bmB\bmB'}{}^\bmC{}_\bmD)
-\bme_{\bmB\bmB'}(\hat{\Gamma}_{\bmA\bmA'}{}^\bmC{}_\bmD) \\
&& \hspace{3cm}-\hat{\Gamma}_{\bmF\bmB'}{}^\bmC{}_\bmD \hat{\Gamma}_{\bmA\bmA'}{}^\bmF{}_\bmB -
\hat{\Gamma}_{\bmB\bmF'}{}^\bmC{}_\bmD
\bar{\hat{\Gamma}}_{\bmA\bmA'}{}^{\bmE'}{}_{\bmB'} +
\hat{\Gamma}_{\bmF\bmA'}{}^\bmC{}_\bmD \hat{\Gamma}_{\bmB\bmB'}{}^\bmF{}_\bmA \\
&& \hspace{3cm}+\hat{\Gamma}_{\bmA\bmF'}{}^\bmC{}_\bmD
\bar{\hat{\Gamma}}_{\bmB\bmB'}{}^{\bmF'}{}_{\bmA'} +
\hat{\Gamma}_{\bmA\bmA'}{}^\bmC{}_{\bmE} \hat{\Gamma}_{\bmB\bmB'}{}^\bmE{}_\bmD
-\hat{\Gamma}_{\bmB\bmB'}{}^\bmC{}_\bmE \hat{\Gamma}_{\bmA\bmA'}{}^\bmE{}_\bmD,\\
&&  \hat{\rho}^{\bmC}{}_{\bmD\bmA\bmA'\bmB\bmB'}\equiv \Xi \phi_{\bmC\bmD\bmA\bmB} \epsilon_{\bmA'\bmB'} + L_{\bmD\bmA'\bmB\bmB'} \epsilon_{\bmD'\bmC'} - L_{\bmD\bmB'\bmA\bmA'} \epsilon_{\bmB'\bmC'}
\end{eqnarray*}
The spinor $f_{\bmA\bmA'}$ is related to the reduced spin connection coefficients via
\[
f_{\bmA\bmA'} = \hat{\Gamma}_{\bmA\bmA'}{}^\bmQ{}_\bmQ.
\]

\medskip
In order to write the Yang-Mills equations in spinorial form it is
observed that because of the symmetries of the gauge fields
$F^\fraka{}_{\bma\bmb}$, its spinorial counterpart
$F^\fraka{}_{\bmA\bmA'\bmB\bmB'}$ has the decomposition
\begin{equation}
F^\fraka{}_{\bmA\bmA'\bmB\bmB'} = \varphi^\fraka{}_{\bmA\bmB} \epsilon_{\bmA'\bmB'} +
\bar{\varphi}^\fraka{}_{\bmA'\bmB'} \epsilon_{\bmA\bmB},
\label{FaradaySpinorDecomposition}
\end{equation}
with $\varphi^\fraka_{\bmA\bmB}=\varphi^\fraka_{(\bmA\bmB)}$. In terms of $\varphi^\fraka_{\bmA\bmB}$ the zero quantities take the form

\begin{eqnarray*}
&& M^\fraka{}_{\bmA\bmA'\bmB\bmB'} \equiv\nabla_{\bmA\bmA'} A^\fraka{}_{\bmB\bmB'} - \nabla_{\bmB\bmB'} A^\fraka{}_{\bmA\bmA'} +
C^\fraka{}_{\frakb\frakc} A^\frakb{}_{\bmA\bmA'} A^\frakc{}_{\bmB\bmB'} -
\varphi^\fraka{}_{\bmA\bmB} \epsilon_{\bmA'\bmB'} -
\bar{\varphi}^\fraka{}_{\bmA'\bmB'} \epsilon_{\bmA\bmB}, \\
&& M^\fraka{}_{\bmA'\bmA} \equiv \nabla^\bmQ{}_{\bmA'} \varphi^\fraka{}_{\bmA\bmQ} + C^\fraka{}_{\frakb\frakc}
A^{\frakb \bmQ}{}_{\bmA'} \varphi^\frakc{}_{\bmA\bmQ}.
\end{eqnarray*}
Furthermore, the spinorial counterpart of the energy-momentum tensor
$T_{\bma\bmb}$ is given by the concise expression
\[
T_{\bmA\bmA'\bmB\bmB'} = \delta_{\fraka\frakb} \varphi^\fraka{}_{\bmA\bmB} \bar{\varphi}^\frakb{}_{\bmA'\bmB'}.
\]
The spinorial counterpart of the rescaled Cotton-York tensor can be
computed from the above expression using the formula 
\begin{eqnarray*}
T_{\bmA\bmB\bmC\bmC'} = 
\tfrac{1}{2} \Xi \nabla_{(\bmA|\bmQ'|}  T_{\bmB)}{}^{\bmQ'}{}_{\bmC\bmC'} 
+ \tfrac{3}{2}\nabla_{(\bmA|\bmQ'|} \Xi T_{\bmB)}{}^{\bmQ'}{}_{\bmC\bmC'} + \nabla^{\bmE\bmE'}\Xi \epsilon_{\bmC(\bmA} T_{\bmB)\bmC'\bmE\bmE'}.
\end{eqnarray*}

\medskip
In terms of the zero-quantities introduced in the previous paragraphs, the
spinorial conformal Einstein-Yang-Mills equations are
given by the conditions
\begin{subequations}
\begin{eqnarray}
&\hat{\Sigma}_{\bmA\bmA'}{}^{\bmC\bmC'}{}_{\bmB\bmB'} \bme_{\bmC\bmC'}=0, \qquad
\hat{\Xi}_{\bmA\bmB\bmC\bmC'\bmD\bmD'}=0 \qquad
\hat{\Delta}_{\bmA\bmA'\bmB\bmB'\bmC\bmC'}=0, \qquad 
\Lambda_{\bmA'\bmA\bmB\bmC}=0,& \label{XCE-YMFE1}\\
& M^\fraka{}_{\bmA\bmA'\bmB\bmB'} =0, \qquad M^\fraka{}_{\bmA'\bmA}=0. & \label{XCE-YMFE2}
\end{eqnarray}
\end{subequations}

\section{General properties of spherically symmetric anti-de
  Sitter-like spacetimes}
\label{Section:GeneralProperties}

For completeness, and to motivate the subsequent discussion, we recall
a basic proposition concerning the behaviour at the conformal boundary of
Einstein-Yang-Mills spacetimes.  In what follows, let
$(\tilde{\mathcal{M}},\tilde{\bmg},\mathfrak{G},\bmF^\fraka,\bmA^\fraka)$ denote an
Einstein-Yang-Mills spacetime ---i.e. a spacetime manifold
$\tilde{\mathcal{M}}$, together with a metric $\tilde{g}_{ab}$ and
$\mathfrak{g}$-valued forms $F^\fraka{}_{ab}$ and $A^\fraka{}_a$ satisfying the Einstein-Yang-Mills
equations \eqref{TensorialEinstein}-\eqref{TensorialYM2}--- and let
$(\mathcal{M},\bmg,\frak{G},\bmF^\fraka,\bmA^\fraka)$ with $\bmg = \Xi^2 \tilde{\bmg}$ denote
a conformal extension of the physical Einstein-Yang-Mills
spacetime. As it is customary, one defines the \emph{conformal boundary}
$\mathscr{I}$ as the set
\[
\mathscr{I} \equiv \{ p\in \mathcal{M} \; | \; \Xi=0   \}.
\]
One then has that: 

\begin{proposition}
For  $\lambda>0$, if the physical energy-momentum tensor of the Yang-Mills field is such
that $T_{\bma\bmb\bmc} = o(\Xi^{-2})$ then $\mathscr{I}$ is a
timelike hypersurface and 
$d_{\bma\bmb\bmc\bmd}=O(\Xi^0)$ at $\mathscr{I}$.
\end{proposition}

The general approach to the proof this results can be found in
e.g. \cite{PenRin86,Ste91}. In view of the above result, in what follows, we will say that an
Einstein-Yang-Mills spacetime
$(\tilde{\mathcal{M}},\tilde{\bmg},\frak{G},\bmF^\fraka,\bmA^\fraka)$ is \emph{anti-de
  Sitter-like} if $\lambda>0$ and there exists a conformal extension
$(\mathcal{M},\bmg,\frak{G},\bmF^\fraka,\bmA^\fraka)$ with $\bmg = \Xi^2 \tilde{\bmg}$ such
that $T_{\bma\bmb\bmc} = o(\Xi^{-2})$.

\subsection{Spherically symmetric anti-de Sitter-like spacetimes}
\label{Section:SphericalSymmetry}
An Einstein-Yang-Mills spacetime
$(\tilde{\mathcal{M}},\tilde{\bmg},\frak{G},\bmF^\fraka,\bmA^\fraka)$
is said to be \emph{spherically symmetric} if the group $SO(3)$ acts
by isometry on $(\tilde{\mathcal{M}},\tilde{\bmg})$ with simply
connected, complete, spacelike 2-dimensional orbits ---see
e.g. \cite{Ehl73}--- and the $\mathfrak{g}$-valued forms $\bmF^\fraka$ and
$\bmA^\fraka$ are invariant under the action of $SO(3)$. Given a
spherically symmetric spacetime, it is natural to introduce the
\emph{quotient manifold} $\tilde{\mathcal{Q}}\equiv
\tilde{\mathcal{M}}/SO(3)$. The manifold $\tilde{\mathcal{Q}}$
inherits from $(\tilde{\mathcal{M}},\tilde{\bmg})$ a 2-dimensional
Lorentzian metric $\tilde{\bmgamma}$ ---the so-called \emph{quotient
metric}. Given a spherically symmetric spacetime, there exists a
function $\tilde{\varrho} : \tilde{\mathcal{Q}} \rightarrow
\mathbb{R}$ such that the physical spacetime metric $\tilde{\bmg}$ can
be written in the \emph{warped product} form
\[
\tilde{\bmg} = \tilde{\bmgamma} - \tilde{\varrho}^2 \bmsigma,
\]
where $\bmsigma$ is the standard metric of $\mathbb{S}^2$. 

\medskip
In what follows, we shall restrict our attention to conformal
extensions of spherically symmetric anti-de Sitter-like
Einstein-Yang-Mills spacetimes
$(\tilde{\mathcal{M}},\tilde{\bmg},\frak{G},\bmF^\fraka,\bmA^\fraka)$
such that the conformal factor $\Xi$ only depends on the coordinates of
the quotient manifold. Under these circumstances, the conformal
extension $(\mathcal{M},\bmg,\frak{G},\bmF^\fraka,\bmA^\fraka)$ is
also spherically symmetric with a quotient manifold $\mathcal{Q}\equiv
\mathcal{M}/SO(3)$ which is a conformal extension of
$\tilde{\mathcal{Q}}$ with $\bmgamma = \Xi^2 \tilde{\bmgamma}$. The
unphysical metric metric $\bmg$ is of the form
\begin{equation}
\bmg = \bmgamma - \varrho^2 \bmsigma, \qquad
\varrho:\mathcal{Q}\rightarrow \mathbb{R}.
\label{SphericallySymmetricUnphysicalMetric}
\end{equation}
Close to the set of points on $\mathcal{Q}$ for which $\Xi=0$ one can
always consider local coordinates $(t,r)$ such that the conformal
boundary $\mathscr{I}$ is described by the condition $r=0$. The normal
to $\mathscr{I}$ is then given by $\mathbf{d}r$ with
\[
\bmg(\mathbf{d}r,\mathbf{d}r) =\bmgamma(\mathbf{d}r,\mathbf{d}r)<0,
\]
and $\bmhS$, the pull-back of $\bmg$ to $\mathscr{I}$, is of the form
\[
\bmhS = A(t) \mathbf{d}t \otimes \mathbf{d}t - B(t) \bmsigma,
\]
with $A(t)$ and $B(t)$ two strictly positive functions such that $B(t)\equiv
\varrho(t,0)$. The 3-dimensional metric $\bmhS$ is a Lorentzian
metric. Without loss of generality one can redefine the coordinate $t$
such that $A(t) = B(t) $ and
\[
\bmhS = A(t) (\mathbf{d}t \otimes \mathbf{d}t - \bmsigma).
\]
This 3-dimensional Lorentzian metric can be readily verified to be
\emph{conformally flat}.

\subsubsection{A symmetry adapted frame}
\label{Section:AdaptedFrame}
The spherical symmetry
of the spacetime can be naturally exploited through the choice of a symmetry adapted frame. Let
$\{\bmX_+,\,\bmX_-\}$ denote a basis of $T\mathbb{S}^2$ consisting of
two linearly independent \emph{complex vectors}. The vectors
$\bmX_+$ and $\bmX_-$ can be chosen so that their duals, $\bmalpha^+$
and $\bmalpha^-$, satisfy the the relations
\begin{subequations}
\begin{eqnarray} &\langle \bmalpha^+, \bmX_+\rangle = 1, \qquad
\langle \bmalpha^-, \bmX_-\rangle=1, \qquad \langle \bmalpha^+, \bmX_-
\rangle=0, \qquad \langle \bmalpha^-,\bmX_+ \rangle
=0,& \label{Basis1}\\ &\bmsigma = 2 (\bmalpha^+ \otimes \bmalpha^- +
\bmalpha^-\otimes \bmalpha^+).& \label{Basis2}
\end{eqnarray}
\end{subequations} Now, let $\{\bmxi_1,\,\bmxi_2,\,\bmxi_3\}$ denote
three (linearly independent) Killing vectors associated to the action
of $SO(3)$ on $\mathcal{M}$. The Lie derivatives $\pounds _{\bmxi_j}
\bmX_\pm$ can be computed from the expressions above ---in particular,
the scaling of
the Killing vectors can always be chosen so that
\[
 \pounds_{\bmxi_j} \bmX_+ = \mbox{i} \bmX_-, \qquad
\pounds_{\bmxi_j}\bmX_- =-\mbox{i} \bmX_+.
\]

\medskip
In what follows, we will consider a basis of $T\mathcal{Q}$ consisting
of two vectors $\{\bme_0,\,\bme_3\}$ which are orthogonal to each
other and normalised with respect
to the metric $\bmgamma$ in such a manner that
\[
\bmgamma(\bme_0,\bme_0)=1, \qquad \bmgamma(\bme_3,\bme_3)=-1,
\]
so that in our signature conventions $\bme_0$ is timelike and $\bme_3$
is spacelike. Letting 
\[
\bmomega^0 \equiv \bmgamma(\bme_0,\cdot), \qquad \bmomega^3 \equiv \bmgamma(\bme_3,\cdot)
\]
one has that
\[
\bmgamma = \bmomega^0 \otimes \bmomega^0 - \bmomega^3 \otimes \bmomega^3.
\]

\medskip
The vectors fields $\{\bme_0,\,\bme_3\}$ on $T\mathcal{Q}$ can be extended to
vectors on the whole of $\mathcal{M}= \mathcal{Q}\times
\mathbb{S}^2$, by requiring that
\[
[\bmxi_i,\bme_0]=0, \qquad
[\bmxi_i,\bme_3]=0,
\]
so that $\bme_0$, $\bme_3$ are vectors on $\mathcal{M}$ which are
invariant under the action of $SO(3)$. The vectors and
$\{\bmX_+,\,\bmX_-\}$ on $T\mathbb{S}^2$ are extended to the rest of
the spacetime by requiring that
\[
[\bme_0, \bmX_\pm]=0, \qquad [\bme_3, \bmX_\pm]=0.
\]
Finally, defining 
\[
\bmm^\flat \equiv \bmg(\bmm,\cdot)= \sqrt{2}\varrho \bmalpha^+, \qquad
\bar{\bmm}^\flat \equiv \bmg(\bar{\bmm},\cdot)= \sqrt{2}\varrho
\bmalpha^-,
\]
and comparing with the metric
\eqref{SphericallySymmetricUnphysicalMetric} one can then write
\begin{eqnarray*}
&& \bmg = \bmomega_0\otimes \bmomega_0 -\bmomega_3 \otimes \bmomega_3 - \bmm^\flat \otimes \bar{\bmm}^\flat - \bar{\bmm}^\flat \otimes \bmm^\flat, \\
&& \phantom{\bmg} = \bmomega_0\otimes \bmomega_0 -\bmomega_3 \otimes \bmomega_3 -
\varrho^2 \bmsigma.
\end{eqnarray*}
In particular, one has that $\pounds_{\bmxi_i}\bmg =0$, and moreover
$\pounds_{\bmxi_i} \bmg^\sharp=0$. Using the Cartan structure
equations one can compute the connection coefficients and components
of the curvature associated to frame basis
$\{\bme_0,\bme_3,\bmm,\bar{\bmm}\}$
\footnote{From the discussion above it follows that this frame can easily be transformed into the typical form of a Newman-Penrose tetrad respecting the spherical symmetry.}. This will not be further
elaborated here. 

\subsubsection{Spherically symmetric Yang-Mills fields}
In what follows, we will be interested in Yang-Mills fields which
inherit the spherical symmetry of the spacetime. Accordingly, we
require that
\begin{equation}
\pounds_{\bmxi_j} \bmF^\fraka =0
\label{SphericalSymmetricGaugeFieldCondition}
\end{equation}
As it can be directly verified using the expressions of the previous
subsection, the most general form of the field strength consistent
with the above requirement can be seen to be given by
\begin{equation}
\bmF^\fraka = F^\fraka_{\bmzero\bmthree} \bmomega^\bmzero \wedge \bmomega^\bmthree +
F^\fraka_{+-} \bmalpha^+ \wedge \bmalpha^-, 
\label{SphericalSymmetricGaugeFieldExpression}
\end{equation}
with
\[
F^\fraka_{\bmzero\bmthree}:\mathcal{Q}\rightarrow \mathbb{R}, \qquad F^\fraka_{+-}:\mathcal{Q}\rightarrow \mathbb{C}.
\]
Requiring $\bmF^\fraka_{\bmzero\bmthree}$ to be real readily implies that $F^\fraka_{+-}$
must be pure imaginary ---that is, one has $F^\fraka_{+-} = - \overline{F^\fraka_{+-}}$. Consistent with the above, we will consider gauge
potentials of the form
\[
\bmA^\fraka= A^\fraka_\bmzero \omega^\bmzero + A^\fraka_\bmthree
\omega^\bmthree + A^\fraka_+ \bmalpha^+ + A^\fraka_- \bmalpha^-,
\]
with 
\[
A^\fraka_\bmzero, \; A^\fraka_\bmthree: \mathcal{Q}\rightarrow
\mathbb{R} \qquad  A^\fraka_+,\; A^\fraka_-:\mathcal{Q}\rightarrow \mathbb{C},
\]
such that $\overline{A^\fraka_+}=A^\fraka_-$ in order to ensure the
reality of $\bmA^\fraka$. Further conditions on $\bmA^\fraka$ can be
obtained from the equation $\pounds_{\bmxi_j} M^\fraka{}_{ab}=0$
taking into account condition
\eqref{SphericalSymmetricGaugeFieldCondition}.These further conditions
will not be required in our subsequent analysis. 

\subsection{The conformal constraint equations at the conformal boundary}
\label{Section:CFEScri}

In order to obtain a deeper understanding about the structural
properties of anti-de Sitter-like spacetimes with matter, it is
convenient to consider the \emph{conformal constraint equations}
---see e.g. \cite{Fri95} for the vacuum case. In what follows let
$\mathcal{S}$ denote a hypersurface (spacelike or timelike) within the
unphysical spacetime $(\mathcal{M},\bmg)$, and let $\bmn$ denote its
normal. Below, the possibility of
$\mathcal{S}$ being spacelike or timelike are discussed simultaneously by  setting $\bmg(\bmn,\bm n)
=\epsilon$, where $\epsilon=\pm 1$. If $\epsilon=1$, then the
hypersurface is spacelike, while if $\epsilon=-1$ then it is
timelike. As in Section
\ref{Section:ConformalEinsteinYangMillsEquations} the various
conformal fields are expressed in terms of their components with
respect to an orthonormal frame $\{ \bme_\bma \}$. If $\mathcal{S}$ is
spacelike then one naturally sets $\bme_\bmzero=\bmn$ and one has
that the frame indices $\bmi,\, \bmj,\,\bmk,\ldots$ take the values
$\bmone,\,\bmtwo,\,\bmthree$. By contrast, in the timelike case one
sets $\bme_\bmthree =\bmn$ and the frame indices $\bmi,\,
\bmj\,\bmk,\ldots$ take the values $\bmzero,\,\bmone,\, \bmtwo$.

\medskip
Now, suppose that the frame $\{ \bme_\bma \}$ has been extended off
$\mathcal{S}$ into $(\mathcal{M},\bmg)$. For either of the cases
$\bme_\bmzero=\bmn$ or $\bme_\bmthree=\bmn$ on $\mathcal{S}$ extend
this notation to $(\mathcal{M},\bmg)$ accordingly. Thus, $\bmn$ is a
vector field in the neighbourhood of $\mathcal{S}$ and we define
\[
\chi_{\bmi\bmj} \equiv \bmg(\nabla_\bmi \bmn, \bme_\bmj) =
\Gamma_\bmi{}^\bmb{}_\perp \eta_{\bmb\bmj} = -\Gamma_\bmi{}^\bmb{}_\bmj \eta_{\bmb\perp}, \qquad \chi \equiv \chi_\bmi{}^\bmi,
\]
where ${}_\perp$ stands for either ${}_\bmzero$ or ${}_\bmthree$. 
Let $\bmh$ denote the metric induced by $\bmg$ on $\mathcal{S}$. The
components of $\bmh$ with respect to the intrinsic frame $\{
\bme_\bmi\}$ will be denoted by $h_{\bmi\bmj} \equiv
\eta_{\bmi\bmj}$. Orthogonal projections of tensors into $\mathcal{S}$
are given by their components with respect to the interior frame $\{
\bme_\bmi\}$. One writes
\[
 \Omega\equiv \Xi|_{\mathcal{S}},\qquad \Sigma\equiv \bmn(\Xi)|_{\mathcal{S}}, \qquad K_{\bmi\bmj} \equiv \chi_{\bmi\bmj}|_{\mathcal{S}},
 \qquad L_\bmi \equiv L_{\bmi\perp}, \qquad d_{\bmi\bmj} \equiv d_{\bmi\perp \bmj \perp},
  \qquad d_{\bmi\bmj\bmk}\equiv d_{\bmi\perp \bmj\bmk}.
\]
The components $d_{\bmi\bmj}$ denote the $\bmn$-electric part
of the Weyl tensor while $d^*_{\bmi\bmj} \equiv -\tfrac{1}{2}
d_{\bmi\bmk\bml} \epsilon_\bmj{}^{\bmk\bml}$ correspond to the
$\bmn$-magnetic part. In the present formalism $K_{\bmi\bmj} $ denotes the second fundamental form of the hypersurface $\mathcal{S}$. 
Note, however, that away from  $\mathcal{S}$ the vector field $\bmn$
need not be hypersurface orthogonal and hence $\chi_{\bmi\bmj} $
cannot be interpreted as the second fundamental form of some
hypersurface in $(\mathcal{M},\bmg)$. Instead $\chi_{\bmi\bmj} $ is a
more general tensor, which is the reason for our distinction in the notation.

In what follows, the Levi-Civita connection of the metric $\bmh$ on
$\mathcal{S}$ will be denoted by $\bmD$. One has that
\[
D_\bmi \bme_\bmj= \Gamma_\bmi{}^\bmk{}_\bmj \bme_\bmk, \qquad \mbox{ on } \mathcal{S},
\]
where $\Gamma_\bmi{}^\bmk{}_\bmj$ are the components intrinsic to
$\mathcal{S}$ of the connection coefficients of the unphysical spacetime
Levi-Civita connection $\bmnabla$ computed using the formula
\[
\nabla_\bma \bme_\bmb = \Gamma_\bma{}^\bmc{}_\bmb \bme_\bmc.
\]

\medskip
In terms of the fields described in the previous paragraphs, the
conformal Einstein constraints at $\mathcal{S}$ are given for
trace-free matter by:
\begin{subequations}
\begin{eqnarray}
&& D_\bmi D_\bmj \Omega = -\epsilon \Sigma K_{\bmi\bmj} -\Omega L_{\bmi\bmj} + s
h_{\bmi\bmj} +\tfrac{1}{2} \Omega^3 T_{\bmi\bmj}, \label{CCEqn1}\\
&& D_\bmi \Sigma =\epsilon K_\bmi{}^\bmk D_\bmk \Omega -\Omega L_\bmi +\tfrac{1}{2} \Omega^3 T_{\bmi\perp}, \\
&& D_\bmi s = -D^\bmk\Omega L_{\bmk\bmi} -\epsilon\Sigma L_\bmi - \tfrac{1}{2} \Omega^2
D^\bmj \Omega T_{\bmi\bmj} + \tfrac{1}{2}\epsilon \Omega^2 \Sigma T_{\perp \bmi}, \label{CCEqn3}\\
&& D_\bmi L_{\bmj\bmk} -D_\bmj L_{\bmi\bmk} = D^\bml \Omega d_{\bml\bmk\bmi\bmj} -\epsilon\Sigma
d_{\bmk\bmi\bmj} -K_{\bmi\bmk} L_\bmj +K_{\bmj\bmk} L_\bmi + \Omega T_{\bmi\bmj\bmk}, \\
&& D_\bmi L_\bmj -D_\bmj L_\bmi = D^\bml \Omega d_{\bml\bmi\bmj} + K_\bmi{}^\bmk L_{\bmj\bmk}
-K_\bmj{}^\bmk L_{\bmi\bmk} + \Omega T_{\bmi\bmj\perp}, \\
&& D^\bmk d_{\bmk\bmi\bmj} =\epsilon \big(K^\bmk{}_\bmi d_{\bmj\bmk}
  -K^\bmk{}_\bmj d_{\bmi\bmk}\big)+  T_{\bmi\bmj\perp}, \\
&& D^\bmi d_{\bmi\bmj}= K^{\bmi\bmk} d_{\bmi\bmj\bmk}+ T_{\perp \bmj \perp}, \\
&& D_\bmj K_{\bmk\bmi} - D_\bmk K_{\bmj\bmi} = \Omega d_{\bmi\bmj\bmk} + h_{\bmi\bmj} L_\bmk -
h_{\bmi\bmk}L_\bmj, \\
&& s_{\bmi\bmj} = \Omega d_{\bmi\bmj} + L_{\bmi\bmj} + \epsilon\big( K (
    K_{\bmi\bmj} -\tfrac{1}{4} K_\bml{}^\bml h_{\bmi\bmj} \big) - K_{\bmk\bmi}
  K_\bmj{}^\bmk + \tfrac{1}{4} K_{\bmk\bml} K^{\bmk\bml} h_{\bmi\bmj}\big), \\
&& \lambda = 6 \Omega s - 3\epsilon \Sigma^2 - 3 D_\bmk \Omega D^\bmk \Omega,
\end{eqnarray}
\end{subequations}
where $s_{\bmi\bmj}$ denotes the components of the Schouten tensor of
the intrinsic metric $\bmh$.

\medskip
As already mentioned in the beginning of this section, the above set-up works both for spacelike $(\epsilon=1)$ and timelike $(\epsilon=-1)$ hypersurfaces. In the following we will need to use both cases as we are considering a spacelike initial hypersurface $\mathcal{S}$ and a timelike conformal boundary $\mathscr{I}$. In order to distinguish the two cases and avoid confusion between the two settings we will adopt the following notational conventions. For the spacelike hypersurface $\mathcal{S}$ we shall adopt the notation as used above. The corresponding notation for the timelike hypersurface $\mathscr{I}$ will be $\bmN=\bme_\bmthree$ for the normal of $\mathscr{I}$, $\SigmaS=\bmN(\Xi)$, 
$\bmhS$ is the intrinsic 3-metric and $\KS_{\bmi\bmj}= (\nabla \bmN)_{\bmi\bmj}\vert_\mathscr{I}$  is the extrinsic curvature of $\mathscr{I}$.

\subsection{The conformal constraints at a timelike conformal boundary}
\label{Subsection:ConstraintsConformalBoundary}

The conformal constraint equations discussed in the previous
paragraphs acquire a particularly simple form at the conformal
boundary of a spacetime. In what follows, it is assumed that both the
components of the energy-momentum tensor $T_{\bma\bmb}$ and of the
rescaled Cotton-York tensor $T_{\bma\bmb\bmc}$ are regular whenever
  $\Xi=0$. Furthermore, it is assumed that $\nabla_a \Xi$ is spacelike
  so that $\epsilon=-1$. As the conformal boundary is a surface of
  constant $\Xi$ with normal $\bmN=\bme_\bmthree$, it follows that the only component of its normal is
  given by $\SigmaS =\bmN (\Xi)$ and, consequently,
  $D_\bmi\Xi= D_\bmi \Omega=0$.

\medskip
Taking into account the observations raised in the previous paragraph,
one has that at a timelike conformal boundary the conformal constraint
equations reduce to:
\begin{eqnarray*}
&& \SigmaS \KS_{\bmi\bmj} =- s
\hS_{\bmi\bmj}, \\
&& D_\bmi \SigmaS =0, \\
&& D_\bmi s = \SigmaS L_\bmi, \\
&& D_\bmi L_{\bmj\bmk} -D_\bmj L_{\bmi\bmk} = \SigmaS
d_{\bmk\bmi\bmj} -\KS_{\bmi\bmk} L_\bmj +\KS_{\bmj\bmk} L_\bmi, \\
&& D_\bmi L_\bmj -D_\bmj L_\bmi =  \KS_\bmi{}^\bmk L_{\bmj\bmk}
-\KS_\bmj{}^\bmk L_{\bmi\bmk}, \\
&& D^\bmk d_{\bmk\bmi\bmj} =-\KS^\bmk{}_\bmi d_{\bmj\bmk}
  + \KS^\bmk{}_\bmj d_{\bmi\bmk}, \\
&& D^\bmi d_{\bmi\bmj}= \KS^{\bmi\bmk} d_{\bmi\bmj\bmk}+ T_{\perp \bmj \perp}, \\
&& D_\bmj \KS_{\bmk\bmi} - D_\bmk \KS_{\bmj\bmi} =  \hS_{\bmi\bmj} L_\bmk -
\hS_{\bmi\bmk}L_\bmj. 
\end{eqnarray*}

In \cite{Fri95} it has been shown that the solution to the above
equations satisfies
\begin{equation}
 \SigmaS = \sqrt{\lambda/3}, \qquad s= \sqrt{\lambda/3}\,\varkappa, \qquad \KS_{\bmi\bmj} = -\varkappa \hS_{\bmi\bmj}, \qquad  L_\bmi = D_\bmi \varkappa, \qquad d^*_{\bmi\bmj} = \sqrt{3/\lambda} k_{\bmi\bmj} 
\label{SolutionConformalConstraintsScri}
\end{equation}
where $\varkappa $ is a smooth, gauge dependent real function $\varkappa:\mathscr{I}\rightarrow \mathbb{R}$ and
\[
k_{\bmi\bmj} \equiv -\tfrac{1}{2} k_{\bmk\bml\bmi} \epsilon_\bmj{}^{\bmk\bml}, \qquad k_{\bmk\bml\bmi}\equiv D_\bmk s_{\bml\bmi}- D_\bml s_{\bmk\bmi},
\]
is the Cotton-York tensor of the 3-metric $\bml$. In this approach
only the components of the electric part of the Weyl tensor need to be
solved for. More precisely, one has the equation
\begin{equation}
D^\bmi d_{\bmi\bmj} = 2\SigmaS T_{\bmj\perp},
\label{DivergenceElectricPartWeyl}
\end{equation}
where it has been used that $T_{\perp \bmj \perp} = 2 \SigmaS
T_{\bmj\perp}$ as a consequence of equation \eqref{RescaledCottonYork}
and the fact that $D_\bmi \SigmaS=0$.

\subsubsection{Conformal gauge transformations at the boundary}
The form of the solution to the conformal constraint equations given
by \eqref{SolutionConformalConstraintsScri} can be simplified
by a suitably choice of the scaling of the unphysical spacetime metric
$\bmg$. Under the transition
\begin{equation}
\bmg \rightarrow \vartheta^2 \bmg, \qquad \Xi \rightarrow \vartheta \Xi,
\label{ConformalGaugeFreedom}
\end{equation}
with $\vartheta\neq 0$ on $\mathcal{M}$ 
one has that at the conformal boundary
\[
{\bmhS} \rightarrow (\vartheta|_{\mathscr{I}})^2 {\bmhS} , \qquad
s|_{\mathscr{I}} \rightarrow (\vartheta^{-1} s + \vartheta^{-2} \nabla^\bma
\Xi \nabla_\bma \vartheta )|_{\mathscr{I}} = (\vartheta^{-1} s + \vartheta^{-2} \SigmaS \bmN( \vartheta) )|_{\mathscr{I}} .
\]
Accordingly, by suitably choosing the values of $\vartheta$ and
$\bmN(\vartheta)$ at $\mathscr{I}$ one can always set $s=0$ at the
conformal boundary. The expressions in 
\eqref{SolutionConformalConstraintsScri} imply that for this
scaling one has $\varkappa=0$ and hence
\[
\KS_{\bmi\bmj}=0, \qquad  L_\bmi=0. 
\]
Thus, in this conformal gauge the conformal boundary is extrinsically
flat and the spatial components of the Schouten tensor $L_{ab}$
coincide with the components of the 3-dimensional (intrinsic) Schouten
tensor of $\mathscr{I}$.

\subsubsection{Spherical symmetry}
In the case of spherically symmetric anti-de Sitter-like spacetimes it
has already been shown that the intrinsic metric of the conformal
boundary is conformally flat, so that
\[
k_{\bmi\bmj}=0, \qquad d^*_{\bmi\bmj}=0 \qquad \mbox{(spherical symmetry)}. 
\]
Thus for setting considered in this article \eqref{SolutionConformalConstraintsScri} can be reduced to $\SigmaS = \sqrt{\lambda/3}$ with the remaining conditions vanishing identically.

\subsection{The Yang-Mills constraints at the conformal boundary}
For completeness it is observed that Yang-Mills equations
\eqref{TensorialYM2}-\eqref{TensorialYM2} also imply constraints on
the conformal boundary. These can be readily seen to be given by :
\begin{subequations}
\begin{eqnarray}
&& D^\bmi E^\fraka{}_\bmi + C^\fraka{}_{\frakb\frakc} A^{\frakb
  \bmi}E^\frakc{}_\bmi=0, \label{YMConstraints1}\\
&& D^\bmi B^\fraka{}_\bmi + C^\fraka{}_{\frakb\frakc} A^{\frakb
  \bmi}B^\frakc{}_\bmi=0, \label{YMConstraints2}
\end{eqnarray}
\end{subequations}
where 
\[
E^\fraka{}_\bmj \equiv F^\fraka{}_{ \perp \bmj}, \qquad
B^\fraka{}_\bmj \equiv F^{\fraka *}{}_{ \perp \bmj},
\]
are the \emph{electric and magnetic parts} of $F^\fraka{}_{\bmi\bmj}$ with
respect to the normal of $\mathscr{I}$ as defined earlier.

\subsection{The mass of anti-de Sitter-like spacetimes}
\label{Section:Mass}
As noted above, the intrinsic metric $\bmhS$ of
$\mathscr{I}$ is conformally flat. Hence, there
exists a timelike conformal Killing vector along $\mathscr{I}$,
i.e. one has $\bmxi \in T \mathscr{I}$ such that
\[
D_{(\bmi} \xi_{\bmj)}  = \tfrac{1}{3}\hS_{\bmi\bmj} D^\bmk \xi_\bmk.
\]
As discussed in \cite{AshMag84},
it is therefore possible make use of the conformal constraint equation
\eqref{DivergenceElectricPartWeyl} to write down an integral balance
equation over a region of the conformal boundary. A direct computation
shows that
\begin{eqnarray*}
&& D^\bmi(d_{\bmi\bmj} \xi^\bmj) = D^\bmi d_{\bmi\bmj} \xi^\bmj +
  d_{\bmi\bmj} D^{(\bmi}\xi^{\bmj)} \\ && \phantom{D^\bmi(d_{\bmi\bmj}
    \xi^\bmj)} = 2 \SigmaS T_{\bmj\perp} \xi^\bmj,
\end{eqnarray*}
where we used \eqref{DivergenceElectricPartWeyl} and the fact that $d_{\bmi\bmj}$ is
$\bmhS$-trace-free. Accordingly, integrating over a region
$\mathscr{R}\subset \mathscr{I}$ bounded by two 2-dimensional surfaces
$\mathscr{C}_1, \; \mathscr{C}_2 \approx \mathbb{S}^2$ and using the
divergence theorem one obtains the balance expression
\begin{equation}
\int_{\mathscr{C}_2} d_{\bmi \bmj} \xi^\bmj \mathbf{d}S^\bmi -
\int_{\mathscr{C}_1} d_{\bmi \bmj} \xi^\bmj \mathbf{d}S^\bmi = 2
\int_{\mathscr{R}} \SigmaS T_{\bmj\perp} \xi^\bmj \mathbf{d}\mu_\bmhS,
\label{FluxScri}
\end{equation}
where $\mathbf{d}\mu_\bmhS$ is the volume element of the 3-metric
$\bmhS$ and the area elements $\mathbf{d}S^\bmi$ are oriented in the direction of the
outward pointing normal. The projection $T_{\bmj\perp}$ is the
so-called \emph{Poynting vector}. A calculation with the
energy-momentum of the Yang-Mills field shows that 
\[
T_{\bmi\perp} = -\epsilon_\bmi{}^{\bmj\bmk} \delta_{\fraka\frakb}
E^\fraka{}_\bmj B^\frakb{}_\bmk.
\]
\emph{In vacuum} the quantity 
\begin{equation}
Q[\bmxi] = \int_{\mathscr{C}} d_{\bmi \bmj} \xi^\bmj \mathbf{d}S^\bmi
\label{Definition:Mass}
\end{equation}
over an arbitrary section $\mathscr{C}$ of $\mathscr{I}$ is
conserved. In particular, if $\bmxi$ is a timelike conformal Killing
vector, then $Q[\bmxi]$ can be interpreted as \emph{the mass of the
anti-de Sitter-like vacuum spacetime}. More generally, in the presence
of a Yang-Mills field, one has that equation \eqref{FluxScri}
describes the change of mass due to the Yang-Mills radiation. The mass
will not change if the Poynting vector vanishes.


\section{Spacetime gauge considerations}
\label{Section:Spacetime gauge considerations}
The purpose of this section is to discuss the gauge
that will be used to obtain an hyperbolic reduction of the conformal
Einstein-Yang-Mills equations. This gauge will be based in the
properties of a class of conformally privileged curves known as
\emph{conformal curves} ---see \cite{LueVal12}--- which, in turn, will
be used to propagate coordinates off an initial hypersurface
$\mathcal{S}$.

\subsection{Conformal curves}
Given a spacetime $(\tilde{\mathcal{M}},\tilde{\bmg})$, a \emph{conformal curve}
is a pair $(\bmx(\tau),\tilde{\bmb}(\tau))$ consisting of a curve
$\bmx(\tau)\in \tilde{\mathcal{M}}$, $\tau\in I\subset \mathbb{R}$
with tangent $\dot{\bmx}(\tau)\in T\tilde{\mathcal{M}}$ and a covector
$\tilde{\bmb}(\tau)\in T^*\tilde{\mathcal{M}}$ along $\bmx(\tau)$ satisfying the
equations
\begin{subequations}
\begin{eqnarray}
&& \tilde{\bmnabla}_{\dot{\bmx}} \dot{\bmx} = -2 \langle \tilde{\bmb},\dot{\bmx} \rangle \dot{\bmx} + \tilde{\bmg}(\dot{\bmx},\dot{\bmx}) \tilde{\bmb}^\sharp, \label{ConformalCurveEquation1}\\
&& \tilde{\nabla}_{\dot{\bmx}} \tilde{\bmb} =\langle \tilde{\bmb},\dot{\bmx}\rangle\tilde{\bmb}-\tfrac{1}{2}\tilde{\bmg}^\sharp(\tilde{\bmb},\tilde{\bmb})\dot{\bmx}^\flat + \tilde{\bmH}(\dot{\bmx},\cdot), \label{ConformalCurveEquation2}
\end{eqnarray}
\end{subequations} 
where $\tilde{\bmH}$ denotes a rank 2 covariant tensor which upon a
conformal rescaling $\bmg =\Xi^2 \tilde{\bmg}$ transforms as:
\[
H_{ab} - \tilde{H}_{ab} = \nabla_a \Upsilon_b + \Upsilon_a \Upsilon_b
- \tfrac{1}{2}g^{cd} \Upsilon_c \Upsilon_d g_{ab}, \qquad \Upsilon_a
\equiv \Xi^{-1} \nabla_a \Xi.
\]
This transformation law is \emph{formally identical to that of the Schouten
tensor}. The conformal curve equations are supplemented by the following
propagation law for a frame $\{ \bme_\bma \}$: 
\begin{equation}
\tilde{\nabla}_{\dot{\bmx}} \bme_\bma = -\langle \tilde{\bmb},\bme_\bma\rangle \dot{\bmx} -\langle \tilde{\bmb},\dot{\bmx}\rangle \bme_\bma + \tilde{\bmg}(\bme_\bma,\dot{\bmx})\tilde{\bmb}^\sharp.
\label{WeylPropagation}
\end{equation}
 If the 1-form $\tilde{\bmb}$ transform as $\bmb= \tilde{\bmb}- \bmUpsilon$
 then it can be verified that 
\begin{subequations}
\begin{eqnarray}
&& \bmnabla_{\dot{\bmx}} \dot{\bmx} = -2 \langle \bmb,\dot{\bmx} \rangle \dot{\bmx} + \bmg(\dot{\bmx},\dot{\bmx}) \bmb^\sharp, \label{ConformalCurveEquationALT1}\\
&& \nabla_{\dot{\bmx}} \bmb =\langle \bmb,\dot{\bmx}\rangle \bmb-\tfrac{1}{2}\bmg^\sharp(\bmb,\bmb)\dot{\bmx}^\flat + \bmH(\dot{\bmx},\cdot), \label{ConformalCurveEquationALT2} \\
&&\nabla_{\dot{\bmx}} \bme_\bma = -\langle {\bmb},\bme_\bma\rangle \dot{\bmx} -\langle {\bmb},\dot{\bmx}\rangle \bme_\bma + {\bmg}(\bme_\bma,\dot{\bmx}){\bmb}^\sharp.
\end{eqnarray}
\end{subequations} 

The tensor $\tilde{\bmH}$ is, in principle, completely
arbitrary. In \cite{LueVal12} it is shown that a convenient choice is
given by
\begin{equation}
\tilde{\bmH}= \tfrac{1}{6}\lambda \tilde{\bmg} \qquad \textmd{i.e.} \qquad \tilde{\bmH} = \tilde{\bmL} - \tfrac{1}{2} \tilde{\bmT}
.
\label{ChoiceH}
\end{equation}

\medskip
\noindent
\textbf{Remark.} A conformal curve is specified by the value of
$\bmx$, $\dot{\bmx}$ and $\tilde{\bmb}$ at some fiduciary time
$\tau_\star$. The corresponding initial values are denoted in the
sequel, respectively, by $\bmx_\star$, $\dot{\bmx}_\star$,
$\tilde{\bmb}_\star$.

\medskip
The following result will be fundamental for the construction of our
gauge ---see \cite{LueVal12} for a proof:

\begin{proposition}
\label{Conformal factor for conformal curves}
Let $(\bmx(\tau),\tilde{\bmb}(\tau))$ denote a timelike solution curve to the conformal curve
equations
\eqref{ConformalCurveEquation1}-\eqref{ConformalCurveEquation2} with
the tensor $\tilde{\bmH}$ given by \eqref{ChoiceH} on a spacetime
$(\tilde{\mathcal{M}},\tilde{\bmg})$. If $\bmg = \Theta^2
\tilde{\bmg}$ is such that
\begin{equation}
\bmg(\dot{\bmx},\dot{\bmx})=1,
\label{ConformalCurveNormalisation}
\end{equation}
then the conformal factor $\Theta$ satisfies 
\begin{equation}
\label{Theta quadratic}
\Theta(\tau) = \Theta_\star + \dot{\Theta}_\star(\tau-\tau_\star) + \tfrac{1}{2}\ddot{\Theta}_\star(\tau-\tau_\star)^2,
\end{equation}
where the coefficients $\Theta_\star \equiv \Theta(\tau_\star)$,
$\dot{\Theta}_\star \equiv \dot{\Theta}(\tau_\star)$ and
$\ddot{\Theta}_\star \equiv \ddot{\Theta}(\tau_\star)$ are constant
along the conformal curve and are subject to the constraints
\[
\dot{\Theta}_\star = \langle \tilde{\bmb}_\star, \dot{\bmx}_\star \rangle\Theta_\star, \qquad \Theta_\star \ddot{\Theta}_\star = \tfrac{1}{2}\bmg^\sharp(\tilde{\bmb}_\star,\tilde{\bmb}_\star) + \tfrac{1}{6}\lambda.
\]
Furthermore, if $\{ \bme_\bma \}_\star$ is an initial
$\bmg$-orthogonal frame with $\bme_{0\star}=\dot{\bmx}_\star$ 
which is subsequently propagated
along the curve $\bmx(\tau)$ according to equation
\eqref{WeylPropagation} then $\{ \bme_\bma \}$ is $\bmg$-orthogonal
for all $\tau$ and along the conformal curve one has that for all
$\tau$
\[
\Theta \tilde{b}_0 = \dot{\Theta}, \qquad \Theta \tilde{b}_\bmi = \Theta_\star \tilde{b}_{\bmi\star} \qquad \mathrm{and} \qquad \langle \bmb,\dot{\bmx} \rangle =0,
\]
where $\tilde{b}_\bmi \equiv \langle \tilde{\bmb},\bme_\bmi \rangle$.
\end{proposition}

\subsection{Conformal curves at the conformal boundary}
In view of the purposes of the present article we are particularly
interested in the behaviour of conformal curves at the conformal
boundary. As it will be seen in the sequel, initial data for the
congruence of conformal curves can be chosen in such a way that a
conformal curve which is initially tangent to $\mathscr{I}$ will
remain tangent to $\mathscr{I}$ for all times.  For our analysis we
will start with a general $\bmg$-orthonormal frame that is adapted to
the conformal boundary in the sense that $\bmN=\bme_3$. It will be
shown that $\bme_3$ is actually Weyl propagated along these boundary
intrinsic conformal curves.  The discussion in this section is
completely general and independent of spherical symmetry.

\medskip
In what follows, it will be convenient to specify a general
orthonormal frame $\{ \bme_\bma\}$ so that $\bme_3$ is normal to
$\mathscr{I}$. This frame can then be extended to a neighbourhood
$\mathcal{U}$ of $\mathscr{I}$ by requiring that $\nabla_{\bme_3}
\bme_3 =0$. It follows from this that 
\[
\Gamma_3{}^\bma{}_\bmb =0 \qquad \mbox{on} \qquad \mathcal{U}.
\]
If one uses \emph{Gaussian coordinates} $(x^\mu)$ on $\mathcal{U}$
based on $\mathscr{I}$ such that $\mathscr{I}=\{p\in \mathcal{U} \,|\,
x^3=0 \}$, it follows then that
\[
e_3{}^\mu = \delta_3{}^\mu, \qquad e_\bma{}^3=\delta_\bma{}^3,
\]
where one has written $\bme_\bma = e_\bma{}^\mu \bmpartial_\mu$. Here
and below $\mu$ refers to components with respect to Gaussian
coordinates $(x^\mu)$, with $\mu = 0, \ldots , 3$. We shall use the
index $\alpha$ to denote the restriction to the coordinate values $0, 1, 2$.

\medskip
The tensorial conformal curve equations
\eqref{ConformalCurveEquationALT1}-\eqref{ConformalCurveEquationALT2} 
can be decomposed in components using the boundary adapted
frame discussed in the previous paragraph. To this end, one writes
\[
\dot{\bmx} =z^\bma \bme_\bma, \qquad \bmb = b_\bma \bmomega^\bma.
\]
Using this decomposition it is not hard to see that the conformal
curve equations split in two groups. Firstly, one has the
\emph{normal equations}:
\begin{eqnarray*}
&& \dot{x}^3 =  z^\bma e_\bma{}^3 = z^3. \\
&& \dot{z}^3 = -\Gamma_\bma{}^3{}_\bmb z^\bma z^\bmb - 2(b_\bmc z^\bmc) z^3 + (z_\bmc z^\bmc) b^3, \\
&& \dot{b}_3 = \Gamma_\bma{}^\bmc{}_3 z^\bma b_\bmc + (b_\bmc z^\bmc)b_3 -
\tfrac{1}{2}(b_\bmc b^\bmc) z_3 + H_{33} z^3 + H_{\bmi3} z^\bmi. 
\end{eqnarray*}
Secondly, for $\bmi=0,1,2$ and $\alpha=0,1,2$  one has the \emph{intrinsic equations}:
\begin{eqnarray*}
&& \dot{x}^\alpha = e_\bma{}^\alpha z^\bma, \\
&& \dot{z}^\bmi = -\Gamma_\bmc{}^\bmi{}_\bmb z^\bmc z^\bmb - 2(b_\bmc z^\bmc) z^\bmi + (z_\bmc z^\bmc)
b^\bmi, \\
&& \dot{b}_\bmi = \Gamma_\bmb{}^\bmc{}_\bmi b_\bmc z^\bmb + (b_\bmc z^\bmc)b_\bmi - \tfrac{1}{2}(b_\bmc
b^\bmc) z_\bmi + H_{3\bmi} z^3 + H_{\bmj\bmi} z^\bmj.
\end{eqnarray*}

\medskip
In order to simplify the analysis of these equations one can
exploit the conformal freedom of the setting and choose an element of
the conformal class of the intrinsic 3-metric $\bmhS$ of $\mathscr{I}$
for which $s=0$. Following the discussion of Section
\ref{Section:CFEScri}, this can always be done locally. Under this
choice of conformal gauge, the solution of the conformal constraint
equations on $\mathscr{I}$ given in
\eqref{SolutionConformalConstraintsScri} implies that
\[
\Gamma_\bma{}^3{}_\bmb =0, \qquad \Gamma_\bma{}^\bmc{}_3 =0, \qquad L_{3\bma}=0.
\]
Now, for the class of conformal curves under consideration, the transformation formula of the tensor $\tilde{\bmH}$ implies that
\[
\bmH = \bmL - \tfrac{1}{2}\Xi^2 \bmT,
\]
where $\bmT$ denotes the \emph{unphysical energy-momentum tensor}. If
the unphysical matter fields are regular at $\mathscr{I}$, it follows
then that
\[
H_{3\bma}=L_{3\bma} =0, \qquad H_{\bmi\bmj}=L_{\bmi\bmj} =s_{\bmi\bmj}.
\]
That is, the (unphysical) 4-dimensional Schouten tensor $\bmL$ is determined by the
3-dimensional Schouten tensor $\bms $ of the intrinsic metric $\bmhS$ of $\mathscr{I}$. From
the previous discussion it follows that the normal subset of the
conformal curve equations reduces to:
\begin{eqnarray*}
&& \dot{x}^3 = z^3, \\
&& \dot{z}^3 = -2 (b_\bmc b^\bmc) z^3 + (z_\bmc z^\bmc) b_3, \\
&& \dot{b}_3 = (b_\bmc z^\bmc) b_3 - \tfrac{1}{2} (b_\bmc b^\bmc) z^3 
.
\end{eqnarray*}
The key observation is that these equations are homogeneous in the unknowns
$(x^3,z^3,b_3)$. Thus, by choosing initial data
\begin{equation}
x^3_\star=0, \qquad \dot{x}^3_\star=0, \qquad b_{3\star}=0,
\label{DataConformalCurvesScri}
\end{equation}
one readily obtains a solution
\[
x^3(\tau)=0, \qquad z^3(\tau)=0, \qquad b_3(\tau)=0
\] 
for later times. Accordingly, conformal curves with initial data given
by \eqref{DataConformalCurvesScri} will remain on
$\mathscr{I}$. Looking now at the intrinsic part of the conformal
curve equations one observes that the equations reduce to
\begin{eqnarray*}
&& \dot{x}^\alpha =  z^\bmi e_\bmi{}^\alpha, \\
&& \dot{z}^\bmi = -\Gamma_\bmk{}^\bmi{}_\bmj z^\bmk z^\bmj - 2(b_\bmj z^\bmj) z^\bmi + (z_\bmj z^\bmj)
b^\bmi, \\
&& \dot{b}_\bmi = \Gamma_\bmj{}^\bmk{}_\bmi z^\bmj b_\bmk + (b_\bmj
z^\bmj)b_\bmi - \tfrac{1}{2}(b_\bmj b^\bmj) z_\bmi + s_{\bmj\bmi} z^\bmj.
\end{eqnarray*}
These equations are the \emph{conformal geodesic equations} for the
conformal structure which is determined by the 3-metric $\bmhS$ on
$\mathscr{I}$.

\medskip
Now let $\bmv$ denote a vector satisfying the Weyl propagation equation
\[
\nabla_{\dot{\bmx}} \bmv = -\langle \bmb,\bmv\rangle\dot{\bmx} -\langle\bmb,\dot{\bmx}\rangle \bmv + \bmg(\bmv,\dot{\bmx}) \bmb^\sharp 
\]
along $\mathscr{I}$. Making the Ansatz $\bmv=\beta \bme_3$, where
$\alpha$ denotes a scalar function on $\mathscr{I}$ one readily finds that $z^3(\tau)=0$ and $b_3(\tau)=0$ imply the equation
\[
\dot{\beta} = -\langle \bmb, \dot{\bmx}\rangle \beta =0.
\]
Thus, $\beta = \beta_\star$ along conformal curves that remain tangent to $\mathscr{I}$.
Accordingly, if one prescribes at some point of the
conformal curve in $\mathscr{I}$ an orthonormal frame $\{\bme_a\}$
containing a vector which is normal to $\mathscr{I}$, one readily
finds that the solution to the Weyl propagation equations will be a
frame along the conformal curve which contains a vector normal to
$\mathscr{I}$. Moreover, since Weyl propagation preserves the
orthogonality of vectors, it follows that the elements of the frame
which are initially tangent to $\mathscr{I}$ will remain so at
later times. In summary, a frame $\{\bme_a\}_\star$ that is initially adapted to the boundary will be Weyl propagated
into a boundary adapted frame $\{\bme_a\}$.

\medskip
The results obtained in the previous paragraphs have been obtained
making use of a particular member of the conformal class $[\bmhS]$. It
is thus of interest to reformulate them in an arbitrary conformal
gauge. To this end one considers on $\mathcal{M}$, a conformal factor
$\vartheta>0$ such that $\vartheta\big|_{\mathscr{I}}=1$ to perform a rescaling as
given in \eqref{ConformalGaugeFreedom}. In this spirit define
\[
\bmg' \equiv \vartheta^2 \bmg = (\Xi')^2 \tilde{\bmg}, \qquad \mathrm{with} \qquad \Xi' =\vartheta \Xi.
\]
This rescaling clearly leaves the boundary
metric $\bmhS$ unchanged in the sense that $\bmhS'
=(\vartheta\big|_{\mathscr{I}})^2 \bmhS $. Furthermore, one finds that
\[
s'\big|_{\mathscr{I}} = (\nabla^a \Xi \nabla_a \vartheta)\big|_{\mathscr{I}} = 
\sqrt{\lambda/3}\, \bme_3 (\vartheta) \big|_{\mathscr{I}},
\]
with $\bmN =\bme_3 $. Comparing the above expression with
\eqref{SolutionConformalConstraintsScri} suggests defining
\[
\varkappa' \equiv \bme_3(\vartheta) |_{\mathscr{I}},
\]
so that one obtains a general-looking solution to the conformal constraint equations on $\mathscr{I}$. Defining a 1-form
\[
\bmk = \vartheta^{-1} \mathbf{d}\vartheta, 
\]
and taking into account the transformation properties of conformal
curves under changes of connections, it follows that
$(\bmx(\tau),\bmb'(\tau))$ with $\bmb'=\bmb-\bmk$ is a solution to the
conformal curve equations for the connection ${\bmnabla}^\prime \equiv
{\bmnabla} + \bmS(\bmk)$. From the definition of $\bmk$ it follows
that ${\bm \nabla}^\prime$ is the Levi-Civita connection of the metric
$\bmg'=\vartheta^2 \bmg$. Notice, in particular, that 
\[
b'_3(\tau)\big |_{\mathscr{I}}= -k_3(\tau)\big|_{\mathscr{I}} = -
\bme_3(\vartheta)|_{\mathscr{I}}  = 
 -\varkappa'. 
\]

\medskip
The discussion of this section can be summarised as follows:

\begin{lemma}
\label{Lemma:ConformalCurvesConformalBoundary}
Let $(\mathcal{M},\bmg) $ be a conformal extension of an anti-de Sitter-like spacetime $(\tilde{\mathcal{M}},\tilde{\bmg}) $ and set $\tilde{\bmH}=\tfrac{1}{6}\lambda
\tilde{\bmg}$. If $\gamma$ is a conformal curve which passes through a
point $p\in \mathscr{I}$, is tangent to $\mathscr{I}$ at $p$ and
satisfies $\langle \bmb, \bmN \rangle|_p = -\varkappa = -\sqrt{3 / \lambda} \, s$ then $\gamma$ remains in
$\mathscr{I}$. Furthermore $\gamma$ defines a conformal geodesic for the conformal
structure of $\mathscr{I}$ and the Weyl propagation
equations in $(\mathcal{M},\bmg) $ admit a solution containing a vector field normal to
$\mathscr{I}$. 
\end{lemma}

This result is the analogue of  Lemma 4.1 in \cite{Fri95} for conformal curves. 

\section{Formulation of an initial boundary value problem}
\label{Section:IBVP}
\begin{figure}
\begin{center}
\includegraphics[scale=1.2]{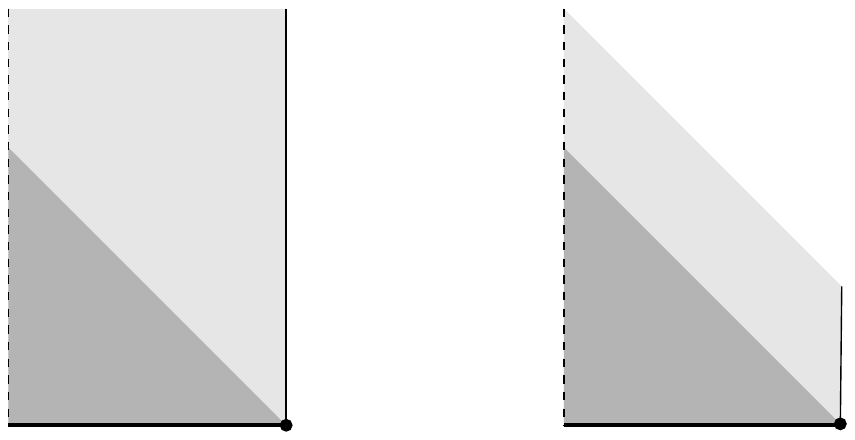}
\put(-5,80){$\mathscr{I}^+$}
\put(-60,-5){$\mathcal{S}$}
\put(-10,-3){$\partial\mathcal{S}$}
\put(-100,30){$D^+(\mathcal{S}\setminus\partial\mathcal{S})$}
\put(-80,90){$D^+(\mathcal{S} \cup \mathscr{I}^+)$}
\end{center}
\caption[Diagram of the construction of anti de Sitter-like spacetimes.]
      {Penrose diagram of the set up for the construction of anti de
        Sitter-like spacetimes as described in the main text. Initial
        data prescribed on
        $\mathcal{S}\setminus\partial\mathcal{S}$ allows to
      recover the dark shaded region
      $D^+(\mathcal{S}\setminus\partial\mathcal{S})$. In
      order to recover $D^+(\mathcal{S} \cup \mathscr{I}^+)$ it
      is necessary to prescribe boundary data on
      $\mathscr{I}^+$. Notice that $D^+(\mathcal{S} \cup \mathscr{I}^+)=J^+(\mathcal{S})$.}
    \label{Figure:AdSDomains}
\end{figure}

Let $(\mathcal{M},\bmg,\bmF^\fraka,\bmA^\fraka,\Xi)$ denote a conformal
extension of an oriented and time oriented spherically symmetric anti-de
Sitter-like spacetime $(\tilde{\mathcal{M}},\tilde{\bmg},\tilde{\bmF}^\fraka,\tilde{\bmA}^\fraka)$ without closed timelike curves.
Furthermore, let $\mathcal{S} \subset \mathcal{M}$ be a smooth, oriented, 
compact, spacelike hypersurface with boundary $\partial\mathcal{S}$ that intersects
the conformal boundary $\mathscr{I}$ so that $\mathcal{S}\cap
\mathscr{I}=\partial \mathcal{S}$. The part of $\mathscr{I}$ in
the future of $\mathcal{S}$ will be denoted by $\mathscr{I}^+$. For
convenience of the discussion, it
will be assumed that the causal future $J^+(\mathcal{S})$
coincides with $D^+(\mathcal{S}\cup \mathscr{I}^+)$, the future domain
of dependence of the set $\mathcal{S}\cup \mathscr{I}^+$ ---a schematic 
depiction of this setting can be seen in Figure \ref{Figure:AdSDomains}.
In the following we will deal with results that are local in time. 
In particular, we will assume that we work with a set
$\mathcal{N}\subset \mathcal{M}$ of the form 
 $\mathcal{N}\approx [0,1]\times\mathcal{S} \subset D^+(\mathcal{S}\cup \mathscr{I}^+)$
 so that $\mathscr{I}^+ \cap \mathcal{N }\approx [0,1]\times \partial
\mathcal{S}$. For the definition of the above causal notions and some
of their properties, see \cite{Wal84}. 


\smallskip
In what follows we will address the
following

\smallskip
\noindent
\textbf{Question:} \emph{What data does one need to specify on 
$\mathcal{S}\cup \mathscr{I}^+$ to reconstruct 
(up to diffeomorphisms) the anti-de Sitter-like Einstein-Yang-Mills spacetime 
$(\tilde{\mathcal{M}},\tilde{\bmg},\tilde{\bmF}^\fraka,\tilde{\bmA}^\fraka)$ in a neighbourhood $U$ of $\mathcal{S}$, where $\mathcal{U}\subset  J^+(\mathcal{S})$
?  }

\medskip 
It is a consequence of the standard Cauchy problem in General Relativity that the solution to
the Einstein-Yang-Mills field equations on the domain of dependence of
$\tilde{\mathcal{S}} \equiv \mathcal{S}\setminus \partial \mathcal{S}$ is
determined in a unique manner, up to diffeomorphisms, by a collection of
tensors $(\tilde{\bmh},\tilde{\bmK},\tilde{\bmE}^\fraka)$ satisfying the 
Einstein-Yang-Mills constraint equations on $\tilde{\mathcal{S}}$. In
order to be able to recover $J^+(\mathcal{S})\setminus
D^+(\tilde{\mathcal{S}})$ one needs to prescribe suitable
initial data on the conformal boundary $\mathscr{I}$. Identifying this
initial data requires a suitable gauge --- that is, a choice of conformal scaling, orthonormal
frame and coordinate system ---  in which
the problem can be analysed. In particular, one needs to specify a gauge near
$\mathscr{I}$. \emph{Following the analysis of the conformal
curves in the previous sections of this article, we will fix the gauge choice using a congruence of conformal curves}.

\subsection{Fixing the gauge}
\label{Section:GaugeConditions}

Following the conventions of Section
\ref{Section:CFEScri}, set $\Omega\equiv
\Xi\big|_{\mathcal{S}}$ below. In order to simplify the subsequent
discussion, it is assumed that the initial spacelike hypersurface $\mathcal{S}$ has 
been chosen so that $\mathcal{S}$ and $\mathscr{I}$ meet orthogonally
---that is, on $\partial \mathcal{S}$ the unit normal $\bmn=\bme_0$ to $\mathcal{S}$ is tangent to $\mathscr{I}$
and one has that $\Sigma = \bmn
(\Xi) = \langle \mathbf{d} \Xi, \bmn \rangle =0$ on $\partial \mathcal{S}$.
Under these circumstances, the conformal factor $\Xi$ can be chosen such that  
\[
\bmn (\Xi) = \Sigma =0, \qquad \mbox{on} \qquad \mathcal{S}.
\]
Moreover, recalling that at the conformal boundary $s$ can be made to
vanish by a convenient choice of conformal gauge, it is assumed that 
\[
s=0, \qquad \mbox{on} \qquad \partial \mathcal{S}.
\]

\medskip
Now, each $p\in \mathcal{S}$ is assumed to be the starting point of
a future directed conformal curve $(\bmx(\tau),\bmb(\tau))$ and an
associated Weyl propagated frame $\{ \bme_\bma \} $. The
parametrisation of the curves is naturally chosen so that $\tau=0$ on
$\mathcal{S}$. For points $p \in \tilde{\mathcal{S}}$ the initial data for
these curves is set in terms of $\tilde{\bmg}$ and its
Levi-Civita connection $\tilde{\bmnabla}$ by the conditions:
\begin{itemize}
\item[(i)] $\dot{\bmx}$ is future directed, orthogonal to
  $\tilde{\mathcal{S}}$ and $\tilde{\bmg}(\dot{\bmx},\dot{\bmx})
  =\Omega^{-2}$,

\item[(ii)] $\bmb_\perp = \Omega^{-1}\mathbf{d}\Omega$ and $\langle
  \bmb,\dot{\bmx} \rangle =0$ (in line with $\Sigma=0$).

\item[(iii)] $\bme_\bmzero =\dot{\bmx}$ and
  $\tilde{\bmg}(\bme_\bma,\bme_\bmb) = \Omega^{-2}\eta_{\bma\bmb}$. 

\end{itemize}

\noindent
\textbf{Remark:} Once the above conditions have been used to fix the
gauge both the set covered by the points of a conformal curve as well
as its parameter $\tau$ are independent of the remaining freedom of
prescribing the frame $\{ \bme_\bma\}$ on $\mathcal{S}$.

\medskip
On suitable neighbourhoods $\mathcal{W} \subset
J^+(\mathcal{S})$ of $\mathcal{S}$ such that their
intersection with conformal curves is connected one has that the curves
$\bmx(\tau)$ define a smooth timelike congruence in $\mathcal{W}$, $\{
\bme_\bma \}$ a smooth frame field and $\bmb$ a smooth 1-form. The
conformal curves can be used to construct a conformal Gaussian coordinate
system $x=(x^\mu)$ on $\mathcal{W}$. For this construction choose a set of local coordinates $(x^\alpha)$ on $\mathcal{S}$ and extend it off $\mathcal{S}$ by keeping $(x^\alpha)$ constant along individual conformal curves. Finally set $x^0=\tau$, the conformal parameter along the conformal curves.

On $\mathcal{W}$ the coefficients $e_\bma{}^\mu =\langle \mathbf{d} x^\mu ,
\bme_\bma\rangle$ of $\bme_\bma$ with respect to the Gaussian
coordinates satisfy $e_\bmzero{}^\mu =
\delta_0{}^\mu$. Notice, however, that in general $e_\bma{}^0=0$ holds only
on  $\mathcal{S}$. The conformal factor $\Xi$ is then fixed on
$\mathcal{W}$ by requiring $\bmg(\bme_\bma,\bme_\bmb)=\eta_{\bma\bmb}$ so
that
\[
\Xi =\Theta.
\]
with $\Theta$ given in \eqref{Theta quadratic}.

In order to include the conformal curves on the conformal boundary 
in the discussion one has to make use of the unphysical metric $\bmg$ and its
Levi-Civita connection $\bmnabla$. In terms of $\bmg$ and
$\bmnabla$,  the conformal curves are
represented by the pair $(\bmx(\tau),\bmf(\tau))$ with $f= \bmb -
\Theta^{-1}\mathbf{d}\Theta$. Accordingly, one has that
\[
\bmf =0, \qquad \mbox{on} \qquad \mathcal{S}.
\]
Now recall that on $\partial \mathcal{S}$ we can fix $b_{3\star}=0$ 
in agreement with our assumption of $s=0$ there. Hence by Lemma
 \ref{Lemma:ConformalCurvesConformalBoundary}
one has that conformal curves which
start on $\partial \mathcal{S}$ remain on $\mathscr{I}$. As $s=0$ on
$\partial \mathcal{S}$ one can write 
\[
s = \Omega \varsigma_\star , \qquad \mbox{on} \qquad \mathcal{S}
\]
where $\varsigma_\star$ is a smooth function on $\mathcal{S}$ with $\varsigma_\star \ne 0$ on $\partial \mathcal{S}$. It follows then that $\Theta_\star=\Omega $, $\dot{\Theta}_\star=0$, $\ddot{\Theta}=\ddot{\Theta}_\star =\Omega \varsigma_\star$ and hence
\begin{equation}
\Theta = \Omega \big( 1- \tfrac{1}{2}\varsigma_\star \tau^2 \big).
\label{ConformalFactorBoundaryAdapted}
\end{equation} 
Moreover, we have
\[
d_0 = \dot{\Theta}, \qquad d_\bmi = d_{\bmi\star} = \bme_\bmi (\Omega)_\star
\equiv (e_\bmi{}^\alpha \partial_\alpha \Omega)_\star
\]
where the functions $\Omega$, $\varsigma_\star$ and
$\bme_\bmi (\Omega)_\star$ defined initially on
$\mathcal{S}$ are extended to $\mathcal{W}$ so that they are constant
along individual conformal curves. The gauge system described in the previous
paragraphs will be known as a \emph{boundary adapted gauge}.  

\subsection{Hyperbolic reduction of the Einstein-Yang-Mills equations}
\label{Subsection:HR_EYMEqns}
The procedure of deducing a symmetric hyperbolic evolution system from 
the extended conformal field equations with the help of a conformal gauge
system based on conformal curves has been discussed in
\cite{LueVal12}. This hyperbolic reduction procedure is conveniently
implemented by means of a space spinor formalism, the details of 
which can be found in \cite{Fri91,Fri98a} and are thus not presented here. 

\medskip
Let $\tau_{AA'}$ denote the spinorial counterpart of the tangent vector to
the conformal curves with normalisation $\tau_{AA'}\tau^{AA'}=2$. In
what follows, a spin dyad $\{\bmepsilon_\bmA \}$ is chosen such that 
\[
\tau_{\bmA\bmA'} \equiv \tau_{AA'} \epsilon_\bmA{}^A
\bar{\epsilon}_{\bmA'}{}^{A'} = \delta_\bmA{}^0 \delta_{\bmA'}{}^{0'} + \delta_\bmA{}^1 \delta_{\bmA'}{}^{1'}. 
\]
The spinor $\tau_{\bmA\bmA'}$ is then used to introduce a space spinor
formalism which allows one to express all the unknowns in the conformal
Einstein-Yang-Mills equations in terms of spinors without primed
indices. In particular, one defines
\begin{eqnarray*}
&\bme_{\bmA\bmB} \equiv  \tau_\bmB{}^{\bmA'} \bme_{\bmA\bmA'}, \qquad
 \hat{\Gamma}_{\bmA\bmB\bmC\bmD} \equiv \tau_\bmB{}^{\bmA'} \hat{\Gamma}_{\bmA\bmA'\bmC\bmD}, \qquad 
f_{\bmA\bmB} \equiv \tau_\bmB{}^{\bmA'} f_{\bmA\bmA'}, & \\
&\hat{L}_{\bmA\bmB\bmC\bmD} \equiv \tau_\bmB{}^{\bmA'} \tau_\bmD{}^{\bmC'} \hat{L}_{\bmA\bmA'\bmC\bmC'} \qquad
 A^\fraka{}_{\bmA\bmB}{} \equiv \tau_\bmB{}^{\bmA'} A^\fraka{}_{\bmA\bmA'}. & 
\end{eqnarray*}
The above fields admit the decomposition \cite{Fri98a} 
\begin{eqnarray*}
 && \bme_{\bmA\bmB} = \tfrac{1}{2}\epsilon_{\bmA\bmB} \bme_\bmQ{}^\bmQ + \bme_{(\bmA\bmB)}, \\
&& \hat{\Gamma}_{\bmA\bmB\bmC\bmD} = \tfrac{1}{\sqrt{2}}(\xi_{\bmA\bmB\bmC\bmD} -\chi_{\bmA\bmB\bmC\bmD}) + \epsilon_{\bmA\bmC} f_{\bmD\bmB},\\
&& f_{\bmA\bmB} =\tfrac{1}{2}\epsilon_{\bmA\bmB} f_\bmQ{}^\bmQ + f_{(\bmA\bmB)}, \\
&& \hat{L}_{\bmA\bmB\bmC\bmD} = \tfrac{1}{2} \epsilon_{\bmA\bmB} \hat{L}_\bmQ{}^\bmQ{}_{\bmC\bmD} + \tfrac{1}{2}\epsilon_{\bmC\bmD} \hat{L}_{\bmA\bmB\bmQ}{}^\bmQ + \hat{L}_{(\bmA\bmB)(\bmC\bmD)}.
\end{eqnarray*}
On the initial hypersurface $\mathcal{S}$, the fields
$\xi_{\bmA\bmB\bmC\bmD}$ and $\chi_{\bmA\bmB\bmC\bmD}$ are associated
to the intrinsic (Levi-Civita) connection and the extrinsic
curvature of $\mathcal{S}$. Since our congruence of conformal curves is not necessarily 
hypersurface orthogonal this interpretation does not hold off $\mathcal{S}$.

\medskip
The gauge conditions associated to the congruence of conformal curves can be expressed in terms of space-spinor objects as
\begin{equation}
\bme_\bmQ{}^\bmQ = \sqrt{2} \bmpartial_\tau, \quad \hat{\Gamma}_\bmQ{}^\bmQ{}_{\bmC\bmD}=0,\quad  f_\bmQ{}^\bmQ=0, \quad \hat{L}_\bmQ{}^\bmQ{}_{\bmC\bmD}= \tfrac{1}{2}\delta_{\fraka\frakb} \phi_{\bmQ\bmC}{}^\fraka \phi^{\dagger\bmQ}{}_\bmD{}^\frakb.
\label{SpinorialGaugeConditions}
\end{equation}
Hence, writing $\bme_{\bmA\bmB} = e_{\bmA\bmB}{}^\mu \partial_\mu$ with $\mu =0,\, 3,\, \pm$ one
has, in particular, that 
\[
\bme_{\bmA\bmB} = \sqrt{2} \epsilon_{\bmA\bmB} \bmpartial_\tau + \big( e^0_{(\bmA\bmB)} \bmpartial_\tau + e^3_{(\bmA\bmB)} \bmpartial_3 + e^+_{(\bmA\bmB)} \bmX_+ + e^-_{(\bmA\bmB)} \bmX_-  \big),
\]
where $\bmX_\pm$ are the vectors on $T\mathbb{S}^2$ introduced in
Section \ref{Section:AdaptedFrame}.

\medskip
In the spirit of the space-spinor formalism, it is convenient to
express the conformal Einstein-Yang-Mills equations
\eqref{XCE-YMFE1}-\eqref{XCE-YMFE2} in terms of the following (equivalent)
space-spinor zero-quantities
\begin{eqnarray*} 
&\hat{\Sigma}_{\bmA\bmB\bmC\bmD} =0, \qquad
\hat{\Xi}_{\bmA\bmB\bmC\bmD\bmE\bmF}=0 \qquad
\hat{\Delta}_{\bmA\bmB\bmC\bmD\bmE\bmF}=0, \qquad 
\Lambda_{\bmA\bmB\bmC\bmD}=0,& \\
& M^\fraka_{\bmA\bmB\bmC\bmD} =0, \qquad M^\fraka_{\bmA\bmB}=0, & 
\end{eqnarray*}
which are obtained by suitably contracting the original spinorial zero
quantities with the spinor $\tau_{\bmA\bmA'}$
---e.g. $\hat{\Sigma}_{\bmA\bmB\bmC\bmD} \equiv \tau_\bmB{}^{\bmA'}
\tau_\bmD{}^{\bmC'} \hat{\Sigma}_{\bmA\bmA'\bmC\bmC'}$. 
Taking into account the gauge conditions \eqref{SpinorialGaugeConditions},
the evolution equations are obtained from
\begin{subequations}
\begin{eqnarray}
& \hat{\Sigma}_\bmQ{}^\bmQ{}_{\bmC\bmD} =0, \qquad
\hat{\Xi}_\bmQ{}^\bmQ{}_{\bmC\bmD\bmE\bmF}=0, \qquad \hat{\Delta}_\bmQ{}^\bmQ{}_{\bmC\bmD\bmE\bmF}=0, \qquad {\Lambda}_{(\bmA\bmB\bmC\bmD)}=0, & \label{SpinorialEvolutionEquations1}\\
& M^\fraka_{(\bmA\bmB)}=0, \qquad M^\fraka_{\bmA\bmQ\bmB}{}^\bmQ +
\epsilon_{\bmA\bmB} (\nabla^{\bmP\bmQ} A^\fraka_{\bmP\bmQ}-\gsF^\fraka)=0,& \label{SpinorialEvolutionEquations2}
\end{eqnarray}
\end{subequations}
where $\gsF^\fraka=\gsF^\fraka(x)$ is an arbitrary gauge source function
expressing the freedom available in the specification of the potential
$A^\fraka_a$ ---see e.g. \cite{Fri85,Fri91}. This gauge
source function allows one to set the divergence of the Yang-Mills potential
equal to any arbitrary function. 

\medskip
\noindent
\textbf{Remark 1.} In vacuum, the conditions
\eqref{SpinorialEvolutionEquations1}-\eqref{SpinorialEvolutionEquations2}
give rise to a symmetric hyperbolic system for the various
(geometric) conformal fields ---see e.g. \cite{Fri98a}. In the
presence of matter this is no longer the case as the Cotton-York
spinor $T_{\bmA\bmB\bmC\bmD}\equiv \tau_\bmD{}^{\bmC'}
T_{\bmA\bmB\bmC\bmC'}$ appearing in the zero-quantities
$\hat{\Delta}_{\bmA\bmB\bmC\bmD\bmE\bmF}$ and
$\Lambda_{\bmA\bmB\bmC\bmD}$ contains derivatives of the spinor field
$\varphi^\fraka_{\bmA\bmB}$ which cannot be eliminated by, say,
using the matter field equations. In order to get around this
complication one introduces the derivatives
$\hat{\nabla}_{\bmA\bmA'} \varphi^\fraka{}_{\bmB\bmC}$ as further
unknowns in the evolution system ---see
\cite{Fri95,LueVal12}. Remarkably, as it will be seen in the sequel,
\emph{under a suitable Ansatz for spherical symmetry all the
non-vanishing derivatives of the field $\varphi^\fraka_{\bmA\bmB}$ can
be obtained from the condition $M^\fraka{}_{\bmA\bmB}=0$}.

\medskip
\noindent
\textbf{Remark 2.} The evolution condition
$\Lambda_{(\bmA\bmB\bmC\bmD)}=0$ leads to the so-called \emph{standard
evolution system} for the independent components of the Weyl spinor
$\phi_{\bmA\bmB\bmC\bmD}$. When considering a boundary value problem, 
the so-called \emph{boundary adapted system} presented in \cite{Fri98a} 
provides a more convenient evolution system. Under a suitable
Ansatz for spherical symmetry (see below) the two evolution systems coincide.

\medskip
In order to show that a solution to the evolution conditions
\eqref{SpinorialEvolutionEquations1}-\eqref{SpinorialEvolutionEquations2}
implies a solution to the conformal Einstein-Yang-Mills equations
\eqref{XCE-YMFE1}-\eqref{XCE-YMFE2}, one constructs a
subsidiary evolution system for the geometric and matter
zero-quantities. In particular, this subsidiary system is homogeneous in the
zero-quantities.
 This lengthy procedure has been discussed in
\cite{Fri91,Fri95,LueVal12}. Due to the homogeneity of the subsidiary system,
the vanishing of the zero-quantities
on the initial hypersurface $\mathcal{S}$ implies that the
zero-quantities must vanish elsewhere. It can be
readily verified that a solution to the full
conformal Einstein-Yang-Mills equations implies a solution to the Einstein-Yang-Mills
equations with anti-de Sitter-like cosmological constant on the set where the conformal
factor is non-vanishing ---see e.g. \cite{Fri85}.  

\subsection{The evolution equations in spherical symmetry}
\label{Section:Evolution equs SS}

The evolution equations \eqref{SpinorialEvolutionEquations1}-\eqref{SpinorialEvolutionEquations2} can be reformulated in terms of scalar equations by decomposing the spinorial fields into irreducible components. Taking into account the symmetries of the fields and the gauge conditions \eqref{SpinorialGaugeConditions} one can write
\begin{eqnarray*}
&& e_{(\bmA\bmB)}{}^\mu = e^\mu_x x_{\bmA\bmB} + e^\mu_y y_{\bmA\bmB} + e^\mu_z z_{\bmA\bmB}, \qquad \mu=0,\;3,\;\pm \\
&& f_{(\bmA\bmB)} = f_x x_{\bmA\bmB} + f_y y_{\bmA\bmB} + f_z z_{\bmA\bmB}, \\
&&  \xi_{\bmA\bmB\bmC\bmD} = \xi_0 \epsilon^0_{\bmA\bmB\bmC\bmD} +
\xi_1 \epsilon^1_{\bmA\bmB\bmC\bmD} +  \xi_2
\epsilon^2_{\bmA\bmB\bmC\bmD} +  \xi_3 \epsilon^3_{\bmA\bmB\bmC\bmD}+
\xi_4 \epsilon^4_{\bmA\bmB\bmC\bmD} \\
&& \hspace{2cm}+ \xi_{\epsilon x} (\epsilon_{\bmA\bmC} x_{\bmB\bmD}+ \epsilon_{\bmB\bmD} x_{\bmA\bmC}) + \xi_{\epsilon y} (\epsilon_{\bmA\bmC} y_{\bmB\bmD}+ \epsilon_{\bmB\bmD} y_{\bmA\bmC})+\xi_{\epsilon z} (\epsilon_{\bmA\bmC} z_{\bmB\bmD} + \epsilon_{\bmA\bmC} z_{\bmB\bmD}),\\
&& \chi_{\bmA\bmB\bmC\bmD} = \chi_0 \epsilon^0_{\bmA\bmB\bmC\bmD} +
\chi_1 \epsilon^1_{\bmA\bmB\bmC\bmD} +  \chi_2
\epsilon^2_{\bmA\bmB\bmC\bmD} +  \chi_3 \epsilon^3_{\bmA\bmB\bmC\bmD}+
\chi_4 \epsilon^4_{\bmA\bmB\bmC\bmD}+ \chi_h h_{\bmA\bmB\bmC\bmD}\\
&& \hspace{2cm}+ \chi_{\epsilon x} (\epsilon_{\bmA\bmC} x_{\bmB\bmD}+ \epsilon_{\bmB\bmD} x_{\bmA\bmC}) + \chi_{\epsilon y} (\epsilon_{\bmA\bmC} y_{\bmB\bmD}+ \epsilon_{\bmB\bmD} y_{\bmA\bmC})+\chi_{\epsilon z} (\epsilon_{\bmA\bmC} z_{\bmB\bmD} + \epsilon_{\bmA\bmC} z_{\bmB\bmD}),\\
&& \hat{L}_{(\bmA\bmB)\bmC\bmD} = \theta_0 \epsilon^0_{\bmA\bmB\bmC\bmD} +  \theta_1 \epsilon^1_{\bmA\bmB\bmC\bmD} +  \theta_2 \epsilon^2_{\bmA\bmB\bmC\bmD} +  \theta_3 \epsilon^3_{\bmA\bmB\bmC\bmD}+  \theta_4 \epsilon^4_{\bmA\bmB\bmC\bmD} + \theta_h h_{\bmA\bmB\bmC\bmD} \\
&& \hspace{3cm} + \theta_{\epsilon y} (\epsilon_{\bmA\bmC} y_{\bmB\bmD}+ \epsilon_{\bmB\bmD} y_{\bmA\bmC})+\theta_{\epsilon z} (\epsilon_{\bmA\bmC} z_{\bmB\bmD} + \epsilon_{\bmA\bmC} z_{\bmB\bmD})\\
&& \hspace{3cm} + \tfrac{1}{\sqrt{2}}\epsilon_{\bmC\bmD}\big( \theta_x x_{\bmA\bmB} + \theta_y y_{\bmA\bmB} + \theta_z z_{\bmA\bmB} \big),  \\
&& \phi_{\bmA\bmB\bmC\bmD}= \phi_0 \epsilon^0_{\bmA\bmB\bmC\bmD} +  \phi_1 \epsilon^1_{\bmA\bmB\bmC\bmD} + \phi_2 \epsilon^2_{\bmA\bmB\bmC\bmD} +  \phi_3 \epsilon^3_{\bmA\bmB\bmC\bmD} +  \phi_4 \epsilon^4_{\bmA\bmB\bmC\bmD}, \\
&& \varphi^\fraka{}_{\bmA\bmB} = \varphi^\fraka_x x_{\bmA\bmB} + \varphi^\fraka_y y_{\bmA\bmB} + \varphi^\fraka_z z_{\bmA\bmB}, \\
&& A^\fraka_{\bmA\bmB} = A^\fraka \epsilon_{\bmA\bmB}  + A^\fraka_x x_{\bmA\bmB} + A^\fraka_y y_{\bmA\bmB} + A^\fraka_z z_{\bmA\bmB},
\end{eqnarray*}
where the basic irreducible spinors
\begin{eqnarray*}
&x_{\bmA\bmB}, \qquad y_{\bmA\bmB}, \qquad z_{\bmA\bmB}, \qquad  h_{\bmA\bmB\bmC\bmD},&\\
& \epsilon^0_{\bmA\bmB\bmC\bmD}, \qquad \epsilon^1_{\bmA\bmB\bmC\bmD},
\qquad \epsilon^2_{\bmA\bmB\bmC\bmD}, \qquad
\epsilon^3_{\bmA\bmB\bmC\bmD}, \qquad \epsilon^4_{\bmA\bmB\bmC\bmD}
\end{eqnarray*}
are defined in the Appendix. For future reference it is convenient to
group the coefficients in the above Ansatz in the vector unknown
\[
 \mathbf{u} \equiv \big(
e^0_x,e^0_y,\cdots\cdots ,A^\fraka_z\big).
\]

\medskip
In the case of a spherically symmetric spacetime ---see the
discussion of Section \ref{Section:SphericalSymmetry}--- one can
introduce a spin dyad adapted to the symmetry of the spacetime.  A suitable Ansatz
for spherical symmetry in the present setting is given by
\begin{subequations}
\begin{eqnarray}
&& e^0_{(\bmA\bmB)} =  e^0_x x_{\bmA\bmB}, \qquad e^3_{\bmA\bmB}
=e^3_x x_{\bmA\bmB}, \qquad e^+_{\bmA\bmB} = e^+_z z_{\bmA\bmB}, \qquad
e^-_{\bmA\bmB} = e^-_y y_{\bmA\bmB}, \label{SSAnsatz1}\\
&& f_{\bmA\bmB} = f_x x_{\bmA\bmB}, \qquad \xi_{\bmA\bmB\bmC\bmD} =
\tfrac{1}{\sqrt{2}}\xi_{\epsilon x} (\epsilon_{\bmA\bmC} x_{\bmB\bmD} +
\epsilon_{\bmB\bmD} x_{\bmA\bmC}), \label{SSAnsatz2}\\
&& \chi_{\bmA\bmB\bmC\bmD} = \chi_2 \epsilon^2_{\bmA\bmB\bmC\bmD} +
\tfrac{1}{3}\chi_h h_{\bmA\bmB\bmC\bmD}, \label{SSAnsatz3}\\
&& \hat{L}_{(\bmA\bmB)\bmC\bmD} = \theta_2 \epsilon^2_{\bmA\bmB\bmC\bmD} +
\tfrac{1}{3}\theta_h h_{\bmA\bmB\bmC\bmD}  + \tfrac{1}{\sqrt{2}}\theta_x\epsilon_{\bmC\bmD}
x_{\bmA\bmB}, \label{SSAnsatz4}\\
&& \phi_{\bmA\bmB\bmC\bmD} = \phi_2 \epsilon^2_{\bmA\bmB\bmC\bmD},
\qquad \varphi^\fraka_{\bmA\bmB} =\varphi^\fraka_x x_{\bmA\bmB}, \label{SSAnsatz5}\\
&& A^\fraka_{\bmA\bmB} = A^\fraka \epsilon_{\bmA\bmB}  + A^\fraka_x
x_{\bmA\bmB} + A_y^\fraka y_{\bmA\bmB} + A^\fraka_z z_{\bmA\bmB} \label{SSAnsatz6}
\end{eqnarray}
\end{subequations}
where all the coefficients in the above Ansatz depend only on the
coordinates of the quotient space $\mathcal{Q}$ and except for $ e^+$, 
$e^-$ and $\varphi^\fraka_x$ are real. Moreover, save for $ e^+$ and
$e^-$ all the coefficients have
spin-weight 0. For more details on the motivations behind the
spherical symmetric Ansatz \eqref{SSAnsatz1}-\eqref{SSAnsatz6}, we
refer the reader to \cite{Fri98a,Val12}. To see that the Ansatz is
consistent with the warp product metric
\eqref{SphericallySymmetricUnphysicalMetric} one can compute $\bmg$ in
terms of the frame coefficients $e^0_x$, $e^1_x$, $e^+_z$, $e^-_y$
using the formula
\[
\bmg = \epsilon_{\bmA\bmB} \epsilon_{\bmA'\bmB'}
\bmomega^{\bmA\bmA'}\otimes 
\bmomega^{\bmB\bmB'},
\]
where $\{\bmomega^{\bmA\bmA'}\}$ denotes the dual cobasis to
$\{\bme_{\bmB\bmB'}\}$ ---that is, one has $\langle
\bmomega^{\bmA\bmA'}, \bme_{\bmB\bmB'}\rangle =\delta_\bmA{}^\bmB
\delta_{\bmA'}{}^{\bmB'}$. 
Defining $\omega^{\bmA\bmB} \equiv
-\bmomega^{\bmA\bmQ'}\tau^\bmB{}_{\bmQ'} $ one has the decomposition
\[
\omega^{\bmA\bmB} = \tfrac{1}{2}\epsilon^{\bmA\bmB} \omega_\bmQ{}^\bmQ
+ \omega^{(\bmA\bmB)}.
\]
Some computations yield the relations 
\begin{eqnarray*}
&&\bmomega_\bmP{}^\bmP = \sqrt{2} \left( \mathbf{d} \tau
  -\frac{e^0_x}{e^1_x} \mathbf{d} x^3 \right), \\
&& {\bmomega}^{(\bmA\bmB)} = - \frac{1}{e^1_x} x^{\bmA\bmB} \mathbf{d} x^3 -
\frac{2}{e^+_z}y^{\bmA\bmB} {\bm\alpha}^+ - \frac{2}{e^-_y} z^{\bmA\bmB} {\bm \alpha}^-.
\end{eqnarray*}
Taking into account the above expressions one further finds that
\[
\bmg = \mathbf{d}\tau \otimes \mathbf{d}\tau -
\frac{e^0_x}{e^1_x}\left( \mathbf{d}\tau \otimes \mathbf{d}x^3 +
  \mathbf{d}x^3 \otimes \mathbf{d}\tau \right) - \left(
  \frac{1}{(e^1_x)^2} -\left( \frac{e^0_x}{e^1_x}\right)^2 \right)
\mathbf{d}x^3 \otimes \mathbf{d}x^3 -\frac{1}{e^+_z
  e^-_y}\bmsigma,
\]
which is in the required warped-product form. An expression for the
gauge field $\bmF^\fraka$ can be computed in a similar manner from 
\[
\bmF^\fraka = F^\fraka_{\bmA\bmA'\bmB\bmB'} \bmomega^{\bmA\bmA'} \otimes \bmomega^{\bmB\bmB'},
\]
which, recalling the split \eqref{FaradaySpinorDecomposition} of the spinorial counterpart of
$F^\fraka_{\bmA\bmA'\bmB\bmB'}$  gives
\[
 F^\fraka_{\bmA\bmA'\bmB\bmB'}=\varphi_x^\fraka x_{\bmA\bmB} \epsilon_{\bmA'\bmB'} +
\bar{\varphi}_x^\fraka \bar{x}_{\bmA'\bmB'} \epsilon_{\bmA\bmB}
\]
so that
\[
\bmF^\fraka = \tfrac{1}{\sqrt{2}}(\varphi^\fraka_x +
\bar{\varphi}^\fraka_x) \bmomega^0\wedge \bmomega^3 + \tfrac{1}{\sqrt{2}} (\varphi^\fraka_x -
\bar{\varphi}^\fraka_x) \bmm^\flat \wedge \bar{\bmm}^\flat.
\]
consistent with the general spherically symmetric expression for the
gauge field in equation
\eqref{SphericalSymmetricGaugeFieldExpression}. A further computation
shows that the associated Hodge dual $\bmF^{\fraka*}$ is given by
\[
\bmF^{\fraka*} = \tfrac{1}{\sqrt{2}}\mbox{i}(\bar{\varphi}^\fraka_x -
\varphi^\fraka_x) \bmomega^0\wedge \bmomega^3 - \tfrac{1}{\sqrt{2}}\mbox{i} (\bar{\varphi}^\fraka_x +
\varphi^\fraka_x) \bmm^\flat \wedge \bar{\bmm}^\flat.
\]
From the above expressions one can read the electric and magnetic
parts of the spherically symmetric gauge field with respect to the
normal of $\mathscr{I}$. One finds that
\[
\bmE^\fraka \big|_{\mathscr{I}}= -\tfrac{1}{\sqrt{2}}(\varphi^\fraka_x +
\bar{\varphi}^\fraka_x) \bmomega^0, \qquad \bmB^\fraka \big|_{\mathscr{I}} =\tfrac{1}{\sqrt{2}}\mbox{i}(\varphi^\fraka_x-\bar{\varphi}^\fraka_x ) \bmomega^0.
\]
Thus, generically, both the electric and magnetic parts of the spherically
symmetric gauge field given by equation
\eqref{SphericalSymmetricGaugeFieldExpression} $\bmF^\fraka$ with
respect to the normal of $\mathscr{I}$ are
non-zero as long as $\varphi_{\bmA\bmB}^\fraka$ is neither real nor
pure imaginary. \emph{Under this assumption, following the discussion of
Section \ref{Section:Mass}, the Poynting vector will be non-vanishing and hence there will be a flow of energy along the conformal boundary. Accordingly,  the mass of the spacetime will not be constant.}

\medskip
Again, for future reference define
\[
\mathbf{u}_{SS} \equiv \big( e^0_x, e^3_x,e^+_z,e^-_y,f_x, \xi_{\epsilon
  x},\chi_2,\chi_h,\theta_2,\theta_h,\theta_x, \phi_2, \varphi^\fraka_x, A^\fraka,A^\fraka_x,A^\fraka_y,A^\fraka_z\big)
\]
The components of the vector unknown $\mathbf{u}$ not appearing in
$\mathbf{u}_{SS}$ will be denoted collectively by
$\mathbf{u}_{NSS}$. \emph{The above Ansatz is consistent with the
evolution equations:} a lengthy computation using the
suite {\tt xAct} for tensorial and spinorial manipulations for {\tt
Mathematica} ---see \cite{GarMar12,xAct}--- shows
that the evolution equations implied by
\eqref{SpinorialEvolutionEquations1}-\eqref{SpinorialEvolutionEquations2}
for the components of $\mathbf{u}_{NSS}$ are homogeneous in
$\mathbf{u}_{NSS}$. Hence they admit the solution
\[
\mathbf{u}_{NSS}=\big(0,\cdots \cdots 0 \big).
\]
Accordingly, components of the spinorial field not appearing in the
Ansatz cannot appear in the course of the evolution if they have been
set initially to zero. Taking into account the gauge conditions
\eqref{SpinorialGaugeConditions},  the evolution equations for the components of
$\mathbf{u}_{SS}$ are given by
\begin{subequations}
\begin{eqnarray}
&& \partial_\tau e_x^0 = \tfrac{1}{3}(\chi_2-\chi_h) e_x^0 -f_x, \label{Reduced:SSEe0}\\
&& \partial_\tau e_x^3 = \tfrac{1}{3}(\chi_2-\chi_h)e_x^3, \label{Reduced:SSEe1}\\
&& \partial_\tau e_z^+ = -\tfrac{1}{6}(\chi_2 +2
\chi_h)e^+_z, \label{Reduced:SSEep} \\
&& \partial_\tau e_y^- = -\tfrac{1}{6}(\chi_2 +2
\chi_h)e^-_y, \label{Reduced:SSEep} \\
&& \partial_\tau f_x = \tfrac{1}{3} (\chi_2 - \chi_h) f_x + \theta_x, \label{Reduced:SSEf}\\
&& \partial_\tau \xi_x = -\tfrac{1}{6}(\chi_2 + 2\chi_h) \xi_x-\tfrac{1}{2}\chi_2 f _x- \theta_x, \label{Reduced:SSExix}\\
&& \partial_\tau \chi_2 = \tfrac{1}{6} (\chi_2 -4\chi_h)\chi_2 -\theta_2 +\Theta\phi_2, \label{Reduced:SSEchi2}\\
&& \partial_\tau \chi_h = -\tfrac{1}{6} \chi^2_2 - \tfrac{1}{3}\chi^2_h -\theta_h -\tfrac{3}{4} \Theta^2 \delta_{\fraka\frakb} \varphi^\fraka_x \bar{\varphi}^\frakb_x, \label{Reduced:SSEchih}\\
&& \partial_\tau \theta_x=\tfrac{1}{3}(\chi_2-\chi_h)\theta_x
-\tfrac{1}{3}d_x \phi_2 +\tfrac{1}{2}\Theta d_x
\delta_{\fraka\frakb}\varphi_x^\fraka \bar{\varphi}_x^\frakb -\tfrac{1}{4} \Theta^2f_x \delta_{\fraka\frakb}\varphi_x^\fraka \bar{\varphi}_x^\frakb, \label{Reduced:SSEthetax}\\
&& \partial_\tau \theta_2 = \tfrac{1}{6}(\chi_2-2\chi_h)\theta_2
-\tfrac{1}{3}\chi_2 \theta_h -
\phi_2 \dot{\Theta}+ \tfrac{1}{4}\Theta^2
\delta_{\fraka\frakb}\varphi_x^\fraka\bar{\varphi}_x^\frakb(3\chi_2+4\chi_h)
\nonumber \\
&& \hspace{2cm} -\Theta
\dot{\Theta}\delta_{\fraka\frakb}
\varphi_x^\fraka\bar{\varphi}_x^\frakb-\tfrac{1}{\sqrt{2}}\Theta^2 \delta_{\frakc\frakd}
\varphi_x^\fraka \bar{\varphi}_x^\frakc C^\frakd{}_{\fraka\frakb} A^\frakb-\tfrac{1}{\sqrt{2}}\Theta^2 \delta_{\frakc\frakd}
\varphi_x^\frakc \bar{\varphi}_x^\fraka \bar{C}^\frakd{}_{\fraka\frakb} A^\frakb,\label{Reduced:SSEtheta2}\\
&& \partial_\tau \theta_h =-\tfrac{1}{6}\chi_2 \theta_2 - \tfrac{1}{3}
\chi_h \theta_h- \Theta \dot{\Theta}\delta_{\fraka\frakb}\varphi_x^\fraka\bar{\varphi}_x^\frakb +\tfrac{1}{4} \Theta^2
\chi_h \delta_{\fraka\frakb}\varphi_x^\fraka
\bar{\varphi}_x^\frakb\nonumber \\
&& \hspace{2cm}-\tfrac{1}{4\sqrt{2}}\Theta^2 \delta_{\fraka\frakd}
\varphi_x^\frakb \bar{\varphi}_x^\fraka C^\frakd{}_{\frakb\frakc} A^\frakc -\tfrac{1}{4\sqrt{2}}\Theta^2 \delta_{\fraka\frakd}
\varphi_x^\fraka \bar{\varphi}_x^\frakb \bar{C}^\frakd{}_{\frakb\frakc} A^\frakc, \label{Reduced:SSEthetah}\\
&& \partial_\tau \phi_2=
-\tfrac{1}{2}(\chi_2+2\chi_h)\phi_2+\tfrac{1}{2}(\chi_2 +2
\chi_h )\Theta \delta_{\fraka\frakb}\varphi_x^\fraka
\bar{\varphi}_x^\frakb
-\dot{\Theta}\delta_{\fraka\frakb}\varphi_x^\fraka\bar{\varphi}_x^\frakb,
\nonumber \\
&& \hspace{2cm} -\tfrac{1}{\sqrt{2}}\Theta \delta_{\frakb\frakd}
\varphi^\fraka_x \bar{\varphi}^\frakb_x C^\frakd{}_{\fraka\frakc}
A^\frakc - \tfrac{1}{\sqrt{2}}\Theta
\delta_{\fraka\frakd}\varphi^\fraka_x \bar{\varphi}^\frakb_x
\bar{C}^\frakd{}_{\frakb\frakc} A^\frakc,\label{Reduced:SSEphi}
\\
&& \partial_\tau \varphi_x^\fraka = -\tfrac{1}{3}\left( \chi_2 +
  2\chi_h\right)\varphi_x^\fraka + \tfrac{1}{\sqrt{2}}C^\fraka{}_{\frakb\frakc}
\varphi_x^\frakb A^\frakc, \label{Reduced:SSEvarphi}\\
&& \partial_\tau A^\fraka -2 e^0_x \partial_\tau A^\fraka_x
-2\sqrt{2} e^3_x \partial_3 A^\fraka_x -\sqrt{2} e^+_z \partial_+
\alpha^\fraka_z -\sqrt{2} e^-_y \partial_- \alpha^\fraka_y \nonumber \\
&&\hspace{1cm}=
4\sqrt{2} \xi_x
A^\fraka_x -2 \sqrt{2} f_x A^\fraka_x -2 \chi_h A^\fraka
+ 2\sqrt{2} F^\fraka(x), \label{Reduced:SSEPotential1} \\
&& \partial_\tau A^\fraka_x - e_x^0 \partial_\tau A^\fraka -
\sqrt{2} e^3_x \partial_3 A^\fraka \nonumber \\
&& \hspace{1cm}=
\tfrac{2}{3}(\chi_2-\chi_h) A^\fraka_x  -\sqrt{2}(\varphi_x^\fraka +\varphi_x^\fraka)
-\sqrt{2} f_x A^\fraka + \sqrt{2} C^\fraka{}_{\frakb\frakc}
A^\frakb_x A^\frakc, \label{Reduced:SSEPotential2}\\
&& \partial_\tau A_y^\fraka - e^+_z \partial_+ A^\fraka =
-\tfrac{1}{3}(\chi_2 +2 \chi_h) A_y^\fraka +
\sqrt{2}C^\fraka{}_{\frakb\frakc} A^\frakb_y
A^\frakc \label{Reduced:SSEPotential3}\\
&& \partial_\tau A^\fraka_z - e^-_y \partial_- A^\fraka =
-\tfrac{1}{3}(\chi_2 +2 \chi_h) A_z^\fraka +
\sqrt{2}C^\fraka{}_{\frakb\frakc} A^\frakb_z A^\frakc. \label{Reduced:SSEPotential4} 
\end{eqnarray}
\end{subequations}

In what follows, the above equations will be called the \emph{spherically
  symmetric conformal evolution equations} for the Einstein-Yang-Mills
system. Note that except for equations
\eqref{Reduced:SSEPotential1}-\eqref{Reduced:SSEPotential4}, the whole
system consists of transport equations along conformal curves. A
quick calculation shows that equations
\eqref{Reduced:SSEPotential1}-\eqref{Reduced:SSEPotential4} can be
rewritten as a symmetric hyperbolic subsystem for the components
$A^\fraka$ and $A^\fraka_x$ of the gauge potentials. 
Equations \eqref{Reduced:SSEthetax}-\eqref{Reduced:SSEtheta2} are 
derived from \eqref{XCFESpinorZQ3} which contains derivatives of the 
energy-momentum tensor and hence of $\varphi^\fraka{}_{\bmB\bmC}$. 
These derivatives obstruct the hyperbolicity of the evolution system. 
In \cite{Fri91} this problem has been dealt with by introducing an 
\emph{auxiliary field} $\psi^\fraka{}_{\bmA\bmA'\bmB\bmC}$ describing the 
components of the derivative $\nabla_{\bmA\bmA'} \varphi^\fraka{}_{\bmB\bmC}$ and their
corresponding evolution equations. This procedure considerably increases the
complexity of the analysis. However, in spherical symmetry the derivative
$\nabla_{\bmA\bmA'} \varphi^\fraka{}_{\bmB\bmC}$ has only 
two independent components ---namely, those
corresponding to the directions $\bme_0$ and $\bme_3$. Moreover, in
the derivation of the evolution equations only the time derivative
$\partial_\tau \varphi^\fraka_x$ appears in the equation. This can be
replaced by an expression not involving derivatives using the
evolution equation \eqref{Reduced:SSEvarphi}. Accordingly, it is
possible to eliminate all terms involving derivatives of the gauge
field spinor $\varphi^\fraka{}_{\bmA\bmB}$ in
\eqref{Reduced:SSEthetax}-\eqref{Reduced:SSEtheta2} and obtain their
present form. Thus, the introduction of the auxiliary term
$\psi^\fraka{}_{\bmA\bmA'\bmB\bmC}$ is not required and the overall
analysis is considerably simplified in the spherically symmetric
setting.

\subsubsection{Initial data for the evolution equations}
The conformal evolution equations
\eqref{Reduced:SSEe0}-\eqref{Reduced:SSEPotential2} need to be
supplemented with spherically symmetric initial data satisfying the conformal constraint
equations on $\mathcal{S}$.  The basic input for this
construction is given by:
\begin{itemize}
 \item[(i)] a \emph{spherically symmetric} conformal factor $\Omega$ on
$\mathcal{S}$ with
\[
\Omega =0, \qquad \mathbf{d}\Omega \neq 0, \qquad \mbox{on} \quad \partial \mathcal{S};
\]

\item[(ii)] 3-dimensional symmetric and \emph{spherically symmetric}
tensors $\bmh$ and $\bmK$ such that 
\[
\tilde{\bmh} = \Omega^{-2} \bmh, \qquad \tilde{\bmK} =\Omega^{-1} \bmK
\]
satisfy, on $\tilde{\mathcal{S}} \equiv
\mathcal{S}\setminus \partial \mathcal{S}$, the
\emph{Hamiltonian} and \emph{momentum constraints}
\[
\tilde{D}^i \tilde{K}_{ij} - \tilde{D}_j \tilde{K} =\tilde{j}_j, \qquad
\tilde{r} - \tilde{K}^2 + \tilde{K}^{ij} \tilde{K}_{ij}
=2(\lambda - \tilde{\rho}),
\]
with $\tilde{r}$ and $\tilde{D}$ denoting, respectively, the Ricci
scalar and Levi-Civita connection of the physical metric
$\tilde{\bmh}$ and $\tilde{\rho}$, $\tilde{j}_j$ the \emph{physical
  energy density} and \emph{energy flux}
of the Yang-Mills field;

\item[(iii)] fields $\varphi^\fraka$ prescribing the initial
  prescription of the gauge field strength which satisfy the Gauss
  constraint implied by the equations $M^\fraka{}_{\bmA'\bmA}=0$,
  which, in the present setting take the from
\[
e^3_x \partial_3 \varphi^\fraka_x + 2 \xi_x \varphi^\fraka_x -
C^\fraka{}_{\frakb\frakc} \varphi^\fraka_x A^\frakc_x =0.
\]

\item[(iv)] an initial prescription of the gauge potential
  $\bmA^\fraka$ satisfying the constraint implied by the equation
  $M^\fraka{}_{\bmA\bmA'\bmB\bmB'}=0$, which, in the present setting
  takes the form
\begin{eqnarray*}
&& e^+_z \partial_+ A^\fraka_z - e^-_y \partial_- A^\fraka_y +
C^\fraka{}_{\frakb\frakc} A^\frakb_y A^\frakc_z =
\sqrt{2}(\bar{\varphi}^\fraka_x - \varphi^\fraka_x), \\
&& e^+_z \partial_+ A^\fraka_x - e^3_x \partial_3 A^\fraka_y - \xi_x
A^\fraka_y -C^\fraka{}_{\frakb\frakc} A^\frakb_x A^\frakc_y=0, \\
&& e^-_y \partial_- A^\fraka_x - e^3_x \partial_3 A^\fraka_z - \xi_x
A^\fraka_z -C^\fraka{}_{\frakb\frakc} A^\frakb_x A^\frakc_z=0.
\end{eqnarray*}

\end{itemize}
Given the above basic data, one can make use the conformal constraints
on $\mathcal{S}$ to compute the required initial data for the
conformal evolution equations using the so-called \emph{conformal
  constraint equations} ---see \cite{Fri83,Fri91} for more
details on this construction. The precise details of this construction
will not be required in the subsequent discussion.
 
\medskip
In addition to the observations made above, one
has that in the present gauge
\[
e^3_x =0, \qquad f_x =0, \qquad \mbox{on} \qquad \partial \mathcal{S}.
\]

\subsection{Identifying the boundary conditions}
\label{Section:boundary conditions}
In order to analyse the possible \emph{maximally dissipative boundary
conditions} associated to the equations
\eqref{Reduced:SSEPotential1}-\eqref{Reduced:SSEPotential2} ---see
e.g. \cite{FriNag99} for an introduction to the underlying theory---
it is convenient to define the new variables
\[
A_+^\fraka \equiv \tfrac{1}{2} (A^\fraka_x+ A^\fraka),
\qquad A^\fraka_-\equiv \tfrac{1}{2}(A^\fraka_x- A^\fraka).
\]

The new variables give rise to the following evolution subsystem
\begin{subequations}
\begin{eqnarray}
&& \big(1 -\tfrac{4}{3}e^0_x\big)\partial_\tau A_+^\fraka +
\tfrac{1}{3}\partial_\tau A^\fraka_- - \tfrac{4\sqrt{2}}{3} e^3_x
\partial_3 A^\fraka_+= \tfrac{4\sqrt{2}}{3} C^\fraka{}_{\frakb\frakc}
A^\frakb_- A^\frakc_++ \tfrac{4\sqrt{2}}{3}(A^\fraka_+
+ A^\fraka_-) \xi_x \nonumber\\
&& \hspace{2cm}  -\tfrac{4\sqrt{2}}{3}
A^\fraka_+ f_x - \tfrac{4\sqrt{2}}{3} \varphi_x^\fraka + \tfrac{2\sqrt{2}}{3}\gsF^\fraka +
\tfrac{2}{9}(A^\fraka_- -5A^\fraka_+)\chi_h  +
\tfrac{4}{9}(A^\fraka_++A^\fraka_-)\chi_2, \label{Reduced:SSEPotential1ALT}\\
&& \partial_\tau A_y^\fraka - e^+_z \partial_+  (A^\fraka_+ - A^\fraka_-) =
-\tfrac{1}{3}(\chi_2 +2 \chi_h) A_y^\fraka +
\sqrt{2}C^\fraka{}_{\frakb\frakc} A^\frakb_y
(A^\frakc_- - A^\frakc_-) \\
&& \partial_\tau A^\fraka_z - e^-_y \partial_- (A^\fraka_+ - A^\fraka_-) =
-\tfrac{1}{3}(\chi_2 +2 \chi_h) A_z^\fraka +
\sqrt{2}C^\fraka{}_{\frakb\frakc} A^\frakb_z (A^\frakc_- - A^\frakc_-) , \\
&& \big(1+ \tfrac{4}{3}e^0_x\big)\partial_\tau A^\fraka_- +
\tfrac{1}{3}\partial_\tau A^\fraka_+ + \tfrac{4\sqrt{2}}{3}
e^3_x \partial_3 A^\fraka_- =
\tfrac{4\sqrt{2}}{3}C^\fraka{}_{\frakb\frakc}
A^\frakb_-A^\frakc_+
-\tfrac{4\sqrt{2}}{3}(A^\fraka_++ A^\fraka_-)\xi_x
 \nonumber \\
&& \hspace{2cm}+\tfrac{4\sqrt{2}}{3}A^\fraka_- f_x -
\tfrac{4\sqrt{2}}{3}\varphi_x^\fraka - \tfrac{2\sqrt{2}}{3}\gsF^\fraka + \tfrac{2}{9}(A^\fraka_+ - 5
A^\fraka_-) \chi_h + \tfrac{4}{9}(A^\fraka_+ +
A^\fraka_-) \chi_2. \label{Reduced:SSEPotential2ALT}
\end{eqnarray}
\end{subequations}
Notice that both equations contain, in their right hand sides, the
gauge source functions $\gsF^\fraka$. The system
\eqref{Reduced:SSEPotential1ALT}-\eqref{Reduced:SSEPotential2ALT} can
be written, schematically, in matrix form as
\[
\mathbb{A}^0 \partial_\tau \mathbf{y} + \mathbb{A}^3 \partial_3
\mathbf{y} = \mathbb{B} \mathbf{y} \qquad \textmd{where} \qquad 
\mathbf{y} = \left(
\begin{array}{c}
A^\fraka_+\\
A^\fraka_y \\
A^\fraka_z \\
A^\fraka_-
\end{array}
\right) .
\]
Key to the identification of maximally dissipative boundary conditions
is the \emph{normal matrix} associated to this evolution subsystem
---that is, the matrix associated to the $\partial_3$ derivative
evaluated at the boundary. A computation shows that it is given by
\[
\mathbb{A}^3|_{\mathscr{I}^+} = \tfrac{4\sqrt{2}}{3}e^3_x 
\left(
\begin{array}{cccc}
-1 & 0 & 0 &0 \\
0 & 0 &0 & 0 \\
0 & 0 & 0& 0 \\
0& 0 &0 & 1
\end{array}
\right).
\]
Maximally dissipative boundary conditions arise from the
identification of the subspaces of $\mathbb{R}^4$ for which the
quadratic form associated to $\mathbb{A}^3|_{\mathscr{I}^+}$ is 
non-positive. Setting for each $\fraka$ (note that here and in the
following no summation is implied in the bundle index $\fraka$)
\[
A^\fraka_- = c^\fraka A^\fraka_+ ,
\]
where $c^\fraka: \mathscr{I} \mapsto \mathbb{R}$ is a smooth functions on $\mathscr{I} $, 
one finds that
\[
(A^\fraka_+,A^\fraka_y, A^\fraka_z, A^\fraka_-) \mathbb{A}^3 
\left(
\begin{array}{c}
A^\fraka_+\\
A^\fraka_y \\
A^\fraka_z \\
A^\fraka_-
\end{array}
\right) = \tfrac{4\sqrt{2}}{3}e^3 \big((c^\fraka)^2-1 \big)\big(A_+^\fraka\big)^2 \leq 0
\qquad \mbox{if and only if} \qquad \vert c^\fraka \vert \leq 1.
\]
Notice, in particular, that it is not possible to prescribe boundary
conditions for the components $A^\fraka_y$ and $A^\fraka_z$. More generally, one
can consider \emph{non-homogeneous  maximally dissipative boundary conditions}
\begin{equation}
A^\fraka_- = c^\fraka A^\fraka_+ + q^\fraka, \qquad
\vert c^\fraka \vert \leq 1,
\label{MaximallyDissipativeBoundaryCondition}
\end{equation}
where $c^\fraka$ and $q^\fraka$ are smooth functions on
$\mathscr{I}$.
A discussion behind the motivation of this procedure to
identify boundary conditions can be found in
e.g. \cite{FriNag99,FriRen00}. 

\medskip
In the following it will be assumed that spherically symmetric
functions $c^\fraka$ and $q^\fraka$ have been specified on
$\mathscr{I}$. As a result their values and derivatives
$\partial^n_\tau c^\fraka$ and $\partial^n_\tau q^\fraka $ for $n \in
\mathbb{N}$ are known along $\mathscr{I}$ and in particular on
$\partial \mathcal{S}$. At this stage, it is important to point out
that there is no guarantee that arbitrary choices of $c^\fraka$ and
$q^\fraka$ give rise to smooth solutions of the field equations. As we
will discuss in the next subsection, smoothness requires that the
initial values for $q^\fraka$, $A^\fraka_+ $ and $A^\fraka_-$ as well
as their derivatives satisfy certain compatibility conditions at
$\partial\mathcal{S}$.

\subsubsection{Corner conditions}
\label{Section:CornerConditions}

In order to obtain smooth solutions to the initial boundary value
problem under consideration, certain compatibility conditions between
the initial data on $\mathcal{S}$ and the boundary conditions
$\mathscr{I}$ need to be satisfied at $\partial
\mathcal{S}$. These conditions are known as \emph{corner
  conditions}.

\medskip
The boundary condition \eqref{MaximallyDissipativeBoundaryCondition}
implies at $\partial \mathcal{S}$ the condition
\[
(q^\fraka)_\circledast =(A^\fraka_-)_\circledast - (c^\fraka)_\circledast (A^\fraka_+)_\circledast 
\]
where the subscript ${}_\circledast$ denotes evaluation at $\partial
\mathcal{S}$. Thus, the value of the function $q^\fraka$ at
$\partial\mathcal{S}$ is fixed by the values of
$(A^\fraka_\pm)_\circledast$ and $(c^\fraka)_\circledast$. If
$q^\fraka=0$, then one sees that $(A^\fraka_+)_\circledast$ and
$(A^\fraka_-)_\circledast$ cannot be prescribed independently of each
other.  The above is the first in a hierarchy of corner conditions.

\medskip
A first order corner condition is obtained by considering the
evolution equations for $A_+^\fraka$ and $A_-^\fraka$ and requiring
consistency with the $\partial_\tau$ derivative of the boundary condition
\eqref{MaximallyDissipativeBoundaryCondition}. Direct
evaluation of the evolution equations
\eqref{Reduced:SSEPotential1ALT}-\eqref{Reduced:SSEPotential2ALT} on
$\partial \mathcal{S}$, using that $e_x^0=0$ and $e_x^3=\sqrt{2}$
there, gives the equations
\begin{subequations}
\begin{eqnarray}
&&\hspace{-1cm} 3(\partial_\tau A_+^\fraka)_\circledast +
(\partial_\tau A^\fraka_-)_\circledast = 8 
(\partial_3 A^\fraka_+)_\circledast+ 4\sqrt{2} C^\fraka{}_{\frakb\frakc}
(A^\frakb_-)_\circledast (A^\frakc_+)_\circledast -
4\sqrt{2}(\varphi^\fraka)_\circledast + 2\sqrt{2}(\gsF^\fraka)_\circledast,  \label{CornerConditions1}\\
&&\hspace{-1cm}  3 (\partial_\tau A^\fraka_-)_\circledast +(\partial_\tau
A^\fraka_+)_\circledast =- 8 (\partial_3 A^\fraka_-)_\circledast 4\sqrt{2}C^\fraka{}_{\frakb\frakc}
(A^\frakb_-)_\circledast(A^\frakc_+)_\circledast -
4\sqrt{2}(\varphi^\fraka)_\circledast - 2\sqrt{2}(\gsF^\fraka)_\circledast. \label{CornerConditions2}
\end{eqnarray}
\end{subequations}
As the $\partial_3$ derivative is intrinsic to the initial
hypersurface, the values of $(\partial_3 A^\fraka_\pm)_\circledast$
can be computed from the initial data. Accordingly, equations
\eqref{CornerConditions1}-\eqref{CornerConditions2} can be read as
as a linear algebraic system for $(\partial_\tau
A_\pm^\fraka)_\circledast$. The $\partial_\tau$-derivative of the boundary condition
\eqref{MaximallyDissipativeBoundaryCondition} then yields 
\[
 (\partial_\tau q^\fraka)_\circledast = (\partial_\tau A^\fraka_-)_\circledast - (c^\fraka)_\circledast (\partial_\tau A_+)_\circledast- (\partial_\tau c^\fraka)_\circledast (A^\fraka_{+\circledast}). 
\]
Consequently, substituting the value of $(\partial_\tau
A_\pm^\fraka)_\circledast$ obtained from solving equations
\eqref{CornerConditions1}-\eqref{CornerConditions2} one obtains an
expression of the form
\[
 (\partial_\tau q^\fraka)_\circledast =
 H\big[(A^\fraka_\pm)_\circledast,(\partial_3 A^\fraka_\pm)_\circledast,(\varphi^\fraka)_\circledast,(F^\fraka)_\circledast,(c^\fraka)_\circledast,(\partial_\tau c^\fraka)_\circledast\big].
\]
Thus, the value of $ (\partial_\tau q^\fraka)_\circledast$ is
completely determined by the restriction of the initial data at
$\partial\mathcal{S}$, the value of the gauge source function
$F^\fraka$ and $c^\fraka$. In the particular case of $q^\fraka=0$, the
above expression should be read as a constraint between $(\partial_3
A^\fraka_+)$ and $(\partial_3 A^\fraka_-)$. 

\medskip
Higher corner conditions can be obtained, as necessary, by considering further
$\partial_\tau$-derivatives of the evolution equations
\eqref{Reduced:SSEPotential1ALT}-\eqref{Reduced:SSEPotential2ALT} and
the boundary condition \eqref{MaximallyDissipativeBoundaryCondition}
and then evaluating these at $\partial \mathcal{S}$. Regarding
derivatives of the form $(\partial_3 \partial^n_\tau A^\fraka_\pm)_\star$
computable from the initial data on $\mathcal{S}$ and the (lower
order) evolution equations, the $n$-th $\partial_\tau$-derivative of equations at $\partial\mathcal{S}$ yields a
linear algebraic system of equations for  $(\partial^{n+1}_\tau
A^\fraka_\pm)_\circledast$. Substituting the result into the $n+1$-th
$\partial_\tau$-derivative of the boundary condition
\eqref{MaximallyDissipativeBoundaryCondition} one obtains the value of
$(\partial^{n+1}_\tau q^\fraka)_\circledast$. Notice that this value
will depend, among other things, on the value of the gauge source
function $F^\fraka$ and its $\partial_\tau$-derivatives at
$\partial\mathcal{S}$. 

\medskip
The procedure described in the previous paragraphs shows that it is
possible to construct, in a neighbourhood of $\partial\mathcal{S}$ in
$\mathscr{I}$ a formal series expansion in $\tau$ for the functions
$q^\fraka$. Thus, the behaviour of the boundary data cannot be
prescribed arbitrarily. In order to ensure smoothness of the 
solution to the boundary value problem in a neighbourhood 
$\mathcal{W}$ of $\partial \mathcal{S}$, the general theory of initial boundary value
problems for symmetric hyperbolic systems requires that the corner conditions
described in the previous paragraphs are satisfied at every order. Notice
that in concrete applications (for example a numerical simulation) it may only be feasible to 
impose the corner conditions to a finite order. The solution so obtained will be
of class $C^k$ for some $k$ rather than $C^\infty$.  

\subsubsection{The frame at the conformal boundary}
The purpose of the present section is to show that the frame
coefficient $e^3_x$, which appears in the normal matrix $\mathbb{A}^3|_{\mathscr{I}^+}$, can be determined purely from initial data on
$\partial \mathcal{S}$. As a result the maximally dissipative
boundary conditions \eqref{MaximallyDissipativeBoundaryCondition} only
prescribe the components $A^\fraka_\pm$ of the gauge
potential.

\medskip
The key to this analysis is the observation that if $\Theta=0$, the evolution
equations \eqref{Reduced:SSEe0}-\eqref{Reduced:SSEPotential2} imply
the following interior subsystem involving the coefficient $e^3_x$:
\begin{subequations}
\begin{eqnarray}
&& \partial_\tau e^3_x = \tfrac{1}{3}(\chi_2-\chi_h)e^3_x, \label{InteriorSubsystem1}\\
&& \partial_\tau \chi_2 = \tfrac{1}{6} (\chi_2 -4\chi_h)\chi_h -\theta_2, \\
&& \partial_\tau \chi_h = -\tfrac{1}{6} \chi^2_2 -
\tfrac{1}{3}\chi^2_h -\theta_h, \\ 
&& \partial_\tau \theta_2 = \tfrac{1}{6}(\chi_2-2\chi_h)\theta_2
-\tfrac{1}{3}\chi_2 \theta_h ,\\
&& \partial_\tau \theta_h =-\tfrac{1}{6}\chi_2 \theta_2 - \tfrac{1}{3}
\chi_h \theta_h. \label{InteriorSubsystem5}
\end{eqnarray}
\end{subequations}
Initial data for the above system is obtained by recalling that in the
present gauge
\[
\Sigma =0, \qquad s=0, \qquad \mbox{on}\qquad \partial\mathcal{S}
\]
so that, in particular,
\begin{equation}
\chi_2=0, \qquad \chi_h =0 \qquad
\mbox{on}\qquad \partial\mathcal{S}.
\label{InteriorSystemData1}
\end{equation}
Now, as $s=\Omega \varsigma_\star$, it follows from the 
conformal constraint equation \eqref{CCEqn3},  using
$\bme_3(\Omega)|_{\partial \mathcal{S}}\neq 0$, that
\[
L_{33}=\varsigma_\star, \qquad \mbox{on}\qquad \partial\mathcal{S}.
\]
As a consequence of the spherical symmetry, the above is the only
non-vanishing component of $L_{33}$. Accordingly, one finds that
\[
L_{(\bmA\bmB)(\bmC\bmD)} = \varsigma_\star x_{\bmA\bmB} x_{\bmC\bmD}.
\]
Thus, using that
\[
x_{(\bmA\bmB)x_(\bmC\bmD)} = 2 \epsilon^2_{\bmA\bmB\bmC\bmD}, \qquad
  h_{\bmA\bmB\bmC\bmD} x^{\bmA\bmB} x^{\bmC\bmD} =1,
\]
one obtains that 
\begin{equation}
\theta_2 = 2\varsigma_\star, \qquad \theta_h = - 
\varsigma_\star \qquad \mbox{on} \qquad \partial\mathcal{S}.
\label{InteriorSystemData2}
\end{equation}
Finally, we observe that $\bme_{01}$ is chosen to give the unit normal $\bmN$ of $\mathscr{I}$ at
$\partial \mathcal{S}$ and that 
\begin{equation}
e^3_x= \sqrt{2}, \qquad \mbox{on} \qquad \partial\mathcal{S}.
\label{InteriorSystemData3}
\end{equation}
Using the initial data \eqref{InteriorSystemData1}, \eqref{InteriorSystemData2} and \eqref{InteriorSystemData3}, it can be
verified that the solution to the interior subsystem \eqref{InteriorSubsystem1}-\eqref{InteriorSubsystem5} is given by
\begin{eqnarray*}
&&e^3_x = \frac{\sqrt{2}}{1+\tfrac{1}{2} \varsigma_\star \tau^2}, \\
&& \chi_2 = - \frac{2\varsigma_\star \tau}{1+\tfrac{1}{2}
  \varsigma_\star \tau^2}, \\
&& \chi_h =  \frac{\varsigma_\star
  \tau}{1+\tfrac{1}{2} \varsigma_\star \tau^2},  \\
&& \theta_2 =\frac{2\varsigma_\star }{1+\tfrac{1}{2} \varsigma_\star
  \tau^2},  \\
&& \theta_h =-\frac{\varsigma_\star }{1+\tfrac{1}{2} \varsigma_\star \tau^2}.
\end{eqnarray*}

The previous analysis can be summarised as follows:

\begin{lemma}
For any solution to the conformal evolution equations
\eqref{Reduced:SSEe0}-\eqref{Reduced:SSEPotential2} satisfying the
conformal constraint equations on $\partial \mathcal{S}$ one
has, irrespective of the values taken by the gauge potential
$A^\fraka{}_{\bmA\bmA'}$ on $\mathscr{I}$, that the normal matrix
is given by 
\[
\mathbb{A}^3|_{\mathscr{I}^+} = \frac{16}{6+3\varsigma_\star \tau^2}
\left(
\begin{array}{cccc}
-1 & 0 & 0 & 0 \\
 0 & 0 & 0 & 0 \\
0 & 0 & 0 & 0 \\
0  & 0 & 0 & 1
\end{array}
\right).
\]
Thus, the maximally dissipative boundary conditions
\eqref{MaximallyDissipativeBoundaryCondition} only imply conditions
on the gauge potentials. 
\end{lemma}

\subsection{Propagation of the constraints}
In order to conclude the construction of solutions to the
conformal Einstein-Yang-Mills field equations, it is necessary to
provide a discussion of the so-called \emph{propagation of the
  constraints}. This analysis requires the construction of a suitable
subsidiary evolution system for the zero-quantities representing the
various conformal field equations. In addition, it is necessary to
consider subsidiary equations for the zero-quantities
\begin{eqnarray*}
&& \delta_\bma \equiv b_\bma - f_\bma - \Theta^{-1} \hat{\nabla}_\bma \Theta,
\\
&& \gamma_{\bma\bmb} \equiv \tfrac{1}{2} \Theta^2 {T}_{\bma\bmb} + \frac{1}{6} \Theta^{-2} \eta_{\bma\bmb}- \hat{L}_{\bma\bmb} -
\hat{\nabla}_\bma b_\bmb - \tfrac{1}{2}S_{\bma\bmb}{}^{\bmc\bmd} b_\bmc
b_\bmd, \\
&& \varsigma_{\bma\bmb} \equiv \hat{L}_{\bma\bmb} -
\hat{\nabla}_{[\bma} f_{\bmb]}, 
\end{eqnarray*}
associated to the conformal gauge used in the
hyperbolic reduction of the conformal field equations.

In order to construct the required subsidiary evolution system we
follow the procedure discussed in \cite{Fri95} for the extended vacuum
conformal field equations and adapt it, as necessary, to the particular features
of the Yang-Mills equations ---see e.g. \cite{Fri91}. A particular
case of this analysis has been carried out in \cite{LueVal12} where
the extended conformal Einstein-Maxwell system was considered. In what
follows, we concentrate on the structural properties of this
computationally intensive analysis. In particular, for the sake of
conciseness, whenever possible we make use of the tensorial
counterpart of the equations. 

\medskip
Assuming that the evolution equations \eqref{SpinorialEvolutionEquations1}-\eqref{SpinorialEvolutionEquations2} hold, a
lengthy computation shows that
\begin{subequations}
\begin{eqnarray}
&& \partial_\tau \delta_\bma = H_\bma[\delta_\bma,\gamma_{\bma\bmb},\varsigma_{\bma\bmb},\hat{\Sigma}_\bma{}^\bmb{}_\bmc], \label{Subsidiary1}\\
&&  \partial_\tau
\gamma_{\bma\bmb} = H_{\bma\bmb}[\gamma_{\bma\bmb}], \label{Subsidiary2}\\
&&  \partial_\tau
\varsigma_{\bma\bmb} = H_{\bma\bmb}[\hat{\Xi}^\bmc{}_{\bmd\bma\bmb}], \label{Subsidiary3}
\end{eqnarray}
\end{subequations}
where the terms $H[\cdots]$ in the right hand side of the equations denote expressions which are homogeneous in
the zero-quantities appearing in brackets. A further computation shows that for the geometric zero-quantities
$\hat{\Sigma}_\bma{}^\bmb{}_\bmc$, $\hat{\Xi}^\bmc{}_{\bmd\bma\bmb}$
and $\hat{\Delta}_{\bmc\bma\bmb}$ one has subsidiary equations of the form
\begin{subequations}
\begin{eqnarray}
&& \partial_\tau \hat{\Sigma}_\bma{}^\bmb{}_\bmc =
H_\bma{}^\bmb{}_\bmc[\hat{\Sigma}_\bma{}^\bmb{}_\bmc,\hat{\Xi}^\bmc{}_{\bmd\bma\bmb}], \label{Subsidiary4}\\
&&  \partial_\tau
\hat{\Xi}^\bmc{}_{\bmd\bma\bmb} = H^\bmc{}_{\bmd\bma\bmb}[\hat{\Sigma}_\bma{}^\bmb{}_\bmc,\hat{\Xi}^\bmc{}_{\bmd\bma\bmb},\hat{\Delta}_{\bmc\bma\bmb},{\Lambda}_{\bmc\bma\bmb}],\label{Subsidiary5}\\
&& \partial_\tau \hat{\Delta}_{\bmc\bma\bmb} =
H_{\bmc\bma\bmb}[\hat{\Sigma}_\bma{}^\bmb{}_\bmc,\hat{\Delta}_{\bmc\bma\bmb},{\Lambda}_{\bmc\bma\bmb},M^\fraka{}_\bma,M^{\fraka
  *}{}_\bma]. \label{Subsidiary6}
\end{eqnarray}
\end{subequations}
For the matter constraint $M^\fraka{}_\bmQ{}^\bmQ$ a direct computation
assuming the evolution equations \eqref{SpinorialEvolutionEquations1}-\eqref{SpinorialEvolutionEquations2}
yields an equation of the form
\begin{equation}
\partial_\tau M^\fraka{}_\bmQ{}^\bmQ =
H[M^\fraka_\bmQ{}^\bmQ,\hat{\Xi}^\bmc{}_{\bmd\bma\bmb}],
\label{SubsidiaryMatter1}
\end{equation}
while the analysis of \cite{Fri91}
shows that the zero-quantity $M^\fraka{}_{\bma\bmb}$ has a subsidiary equation of the form
\begin{equation}
\partial_\tau M^\fraka{}_{\bma\bmb} = H[M^\fraka{}_{\bma\bmb},
M^\fraka{}_\bma, M^{\fraka *}{}_\bma,
\hat{\Sigma}_\bma{}^\bmb{}_\bmc,\hat{\Xi}^\bmc{}_{\bmd\bma\bmb}]
\label{SubsidiaryMatter2}.
\end{equation}
Finally, for $Q^\fraka \equiv \nabla^{\bmP\bmQ}
A^\fraka_{\bmP\bmQ}-F^\fraka$, one finds a subsidiary system of the
form
\begin{equation}
\partial_\tau Q^\fraka = H^\fraka[Q^\fraka,M^\fraka{}_{\bma\bmb},
M^\fraka{}_\bma, M^{\fraka *}{}_\bma,
\hat{\Sigma}_\bma{}^\bmb{}_\bmc,\hat{\Xi}^\bmc{}_{\bmd\bma\bmb}]  
\label{SubsidiaryMatter3}
\end{equation}

Besides their homogeneity in the zero-quantities, the key feature of
the subsidiary equations \eqref{Subsidiary1}-\eqref{Subsidiary3},
\eqref{Subsidiary4}-\eqref{Subsidiary6}, \eqref{SubsidiaryMatter1},
\eqref{SubsidiaryMatter2} and \eqref{SubsidiaryMatter3} is that they
are all transport equations. Accordingly, they do not have to be
supplemented by a boundary condition.

The analysis of the subsidiary equation associated to the
zero-quantity $\Lambda_{\bmc\bma\bmb}$ is much more
delicate. Following the strategy discussed in \cite{Fri95}, the
boundary adapted Bianchi evolution system implies a subsidiary
equation system of the form
\begin{subequations}
\begin{eqnarray}
&& \partial_\tau C_{\bmzero\bmzero} + e^\mu_{\bmzero\bmzero}\partial_\mu
C_{\bmzero\bmone} = U_{\bmzero\bmzero}[\hat{\Sigma}_\bma{}^\bmb{}_\bmc,\hat{\Xi}^\bmc{}_{\bmd\bma\bmb},\varsigma_{\bma\bmb},M^\fraka{}_\bma,M^{\fraka
  *}{}_\bma], \label{SubsidiaryBianchi1}\\
&& \partial_\tau C_{\bmzero\bmone} +
e^\mu_{\bmzero\bmzero}  \partial_\mu C_{\bmone\bmone} -
e^\mu_{\bmone\bmone} \partial_\mu C_{\bmzero\bmzero} =
U_{\bmzero\bmone}[\hat{\Sigma}_\bma{}^\bmb{}_\bmc,\hat{\Xi}^\bmc{}_{\bmd\bma\bmb},\varsigma_{\bma\bmb},M^\fraka{}_\bma,M^{\fraka
  *}{}_\bma], \label{SubsidiaryBianchi2}\\
&& \partial_\tau C_{\bmone\bmone} - e^\mu_{\bmone\bmone} \partial_\mu
C_{\bmzero\bmone} = U_{\bmone\bmone}[\hat{\Sigma}_\bma{}^\bmb{}_\bmc,\hat{\Xi}^\bmc{}_{\bmd\bma\bmb},\varsigma_{\bma\bmb},M^\fraka{}_\bma,M^{\fraka
  *}{}_\bma], \label{SubsidiaryBianchi3}
\end{eqnarray}
\end{subequations}
for the components $C_{\bmA\bmB}$ of the Bianchi constraints. It can
be readily verified that the above system has a vanishing normal
matrix. Accordingly, the subsidiary equations
\eqref{SubsidiaryBianchi1}-\eqref{SubsidiaryBianchi3} do not give rise
to boundary conditions.

\medskip
From the homogeneity in the zero quantities of the subsidiary equations
\eqref{Subsidiary1}-\eqref{Subsidiary3},
\eqref{Subsidiary4}-\eqref{Subsidiary6},
\eqref{SubsidiaryMatter1}, \eqref{SubsidiaryMatter2} and
\eqref{SubsidiaryBianchi1}-\eqref{SubsidiaryBianchi3}, 
and the absence of further boundary conditions we readily
obtain the following \emph{reduction lemma}:

\begin{lemma}[Reduction Lemma]
\label{ReductionLemma}
Let $p\in \partial \mathcal{S}$, where $\mathcal{U}$ is an open
neighbourhood of $p$ in $[0,\infty)\times \mathcal{S}$ and
$\mathcal{V}\equiv \mathcal{U}\cap(\mathcal{S} \cup
\mathscr{I})$. Assume one has a smooth solution to the conformal evolution equations
\eqref{SpinorialEvolutionEquations1}-\eqref{SpinorialEvolutionEquations2}
for data on $\mathcal{V}$ in the boundary adapted gauge which satisfy
on $\mathcal{V}\cap \mathcal{S}$ the conformal constraint
equations. Finally, denote by $\bmg$ the metric obtained from the
orthonormal frame $\bme_\bma$ and by $D^+(\mathcal{V})$ the future
domain of dependence of $\mathcal{V}$ in $\mathcal{U}$ with respect to
$\bmg$. Then the extended conformal Einstein-Yang-Mills equations
are are satisfied on $D^+(\mathcal{V})$ by the conformal fields
solving the conformal evolution equations
\eqref{SpinorialEvolutionEquations1}-\eqref{SpinorialEvolutionEquations2}.
\end{lemma}

\section{Main result}
\label{Section:Main result}

Given a hypersurface $\mathcal{S}$ with boundary $\partial
\mathcal{S}$, the natural domain to look for solutions to an
initial boundary value problem for the conformal Einstein-Yang-Mills
equations \eqref{Reduced:SSEe0}-\eqref{Reduced:SSEPotential2} with
initial data prescribed on $\mathcal{S}$ and boundary conditions
on $[0,\infty)\times \partial \mathcal{S}$ is of the form
$[0,\infty)\times \mathcal{S}$. Using the theory of initial
boundary value problems for symmetric hyperbolic systems with
maximally dissipative boundary conditions as described in
\cite{Gue90,Rau85} one has the following existence theorem for
solutions to the conformal Einstein-Yang-Mills system:

\begin{theorem}
\label{Theorem:MainTheoremPreciseVersion}
Given spherically symmetric initial data $\mathbf{u}_\star$ for the conformal
Einstein-Yang-Mills field
equations with an anti-de Sitter-like cosmological constant on an
initial hypersurface $\mathcal{S}$, smooth gauge source functions $\gsF^\fraka$ on
$[0,\infty)\times \mathcal{S}$, and smooth functions $c^\fraka$ and
$q^\fraka$ on $[0,\infty)\times \partial \mathcal{S}$ with $\vert c^\fraka \vert \leq 1$
satisfying the required
corner conditions to any order, there exists for some $T>0$ a
unique solution $\mathbf{u}$ of the conformal field equations on a
domain
\[
\mathcal{M}_T \equiv \{ p\in [0,\infty)\times \mathcal{S} \; | \;
0\leq \tau(p) \leq T \}
\]
such that
\[
\mathbf{u}|_\mathcal{S} = \mathbf{u}_\star, \qquad \qquad \qquad
(A^\fraka_- - c^\fraka A^\fraka_+)|_{\mathscr{I}\cap\mathcal{M}_T} = q^\fraka,
\]
and
\[
\nabla^a A^\fraka{}_a =\gsF^\fraka \quad \mbox{on} \quad \mathcal{M}_T.
\]
Moreover, the fields $(\tilde{\bmg},\bmF^\fraka,\bmA^\fraka)$ obtained
from the solution $\mbfu$ to the conformal Einstein-Yang-Mills
equations constitute a solution to the Einstein-Yang-Mills system on
$\tilde{\mathcal{M}}_T\equiv \mathcal{M}_T\setminus \mathscr{I}$ for
which $\mathscr{I}$ represents null infinity. 
\end{theorem}

The above result constitutes an alternative formulation of the main
theorem stated in the introductory section. 

\begin{proof}
The existence of solutions to the conformal evolution system
 \eqref{Reduced:SSEe0}-\eqref{Reduced:SSEPotential2} follows from the assumptions on the initial data,
boundary data and the corner conditions using the general theory of initial
boundary value problems with maximally dissipative boundary data as
given in \cite{Gue90,Rau85} and applied in \cite{Fri95,FriNag99}. Once
a solution to the reduced system is obtained, a solution to the full
conformal Einstein-Yang-Mills equations follows by the assumption on
the initial data and the \emph{Reduction Lemma}. From here, general properties of the conformal
Einstein field equations imply the existence of a solution to the
Einstein-Yang-Mills equations away from the conformal boundary ---see
\cite{Fri83}. 
\end{proof}

\section{Concluding remarks}
\label{Section:Discussion}
The purpose of this article has been the formulation of an
initial boundary value problem for the Einstein-Yang-Mills equations
in a conformal setting which allows to show the local existence of a
big class spherically symmetric solutions to these equations which
behave, asymptotically, like the anti-de Sitter spacetime. The use of
conformal methods allows to identify a great class of boundary
conditions ensuring the well-posedness of the problem. 

The present analysis is a first natural step towards a 
formulation of the local existence of non-symmetric anti-de Sitter like solutions to the
Einstein-Yang-Mills equations. As a consequence of the spherical
symmetry, no boundary data involving the geometric variables can be
prescribed. This situation is bound to change in the general,
non-symmetric, setting  where intuitively one would expect to be able
to prescribe a linear combination of components of the Weyl tensor
(expressed in terms of a boundary adapted frame). On the matter side,
in addition to the boundary conditions for the gauge potential 1-form,
boundary conditions for the gauge field will be required. The details
of this intricate construction will be discussed elsewhere.

We expect the spherically symmetric conformal evolution system
\eqref{Reduced:SSEe0}-\eqref{Reduced:SSEPotential2} to be amenable to a numerical implementation. The
simulations obtained from such implementation will give valuable
information concerning the global existence and stability of the local
solutions constructed in the present work.

\section*{Acknowledgements}
CL is pleased to acknowledged financial support from the grant CERN/FP/123609/2011 and a JSPS fellowship. CL would like to thank UCL for a Visiting research fellowship and UCL and Queen Mary for their hospitality during key parts of this research.
\section*{Appendix}

Any symmetric rank 2 spinorial field $X_{\bmA\bmB} = X_{(\bmA\bmB)}$ can be
decomposed in terms of the \emph{basic spinors}
\[
x_{\bmA\bmB} \equiv \sqrt{2} \delta_{(\bmA}{}^0 \delta_{\bmB)}{}^1, \qquad y_{\bmA\bmB} \equiv -\tfrac{1}{\sqrt{2}} \delta_\bmA{}^1 \delta_\bmB{}^1, \qquad z_{\bmA\bmB}\equiv \tfrac{1}{\sqrt{2}}\delta_\bmA{}^0 \delta_\bmB{}^0.
\]
From the above, only $x_{\bmA\bmB}$ has spin-weight $0$. Similarly,
any totally symmetric rank 4 spinorial field $X_{\bmA\bmB\bmC\bmB}$
can be constructed using the basic spinors 
\begin{eqnarray*}
&\epsilon^0_{\bmA\bmB\bmC\bmD} \equiv \delta_{(\bmA}{}^0\delta_\bmB{}^0
\delta_\bmC{}^0 \delta_{\bmD)}{}^0, \qquad
\epsilon^1_{\bmA\bmB\bmC\bmD} \equiv \delta_{(\bmA}{}^0\delta_\bmB{}^0
\delta_\bmC{}^0 \delta_{\bmD)}{}^1, \qquad 
\epsilon^2_{\bmA\bmB\bmC\bmD} \equiv \delta_{(\bmA}{}^0\delta_\bmB{}^0
\delta_\bmC{}^1 \delta_{\bmD)}{}^1& \\
&\epsilon^3_{\bmA\bmB\bmC\bmD} \equiv \delta_{(\bmA}{}^0\delta_\bmB{}^1
\delta_\bmC{}^1 \delta_{\bmD)}{}^1, \qquad \epsilon^4_{\bmA\bmB\bmC\bmD} \equiv \delta_{(\bmA}{}^1\delta_\bmB{}^1
\delta_\bmC{}^1 \delta_{\bmD)}{}^1.&
\end{eqnarray*}
More general rank 4 spinors with the pairwise symmetry
$X_{\bmA\bmB\bmC\bmD}=X_{(\bmA\bmB)(\bmC\bmD)}$ are constructed using
the above and the combinations
\begin{eqnarray*}
& x_{\bmA\bmC} \epsilon_{\bmB\bmD} + x_{\bmB\bmD} \epsilon_{\bmA\bmC},
\qquad y_{\bmA\bmC} \epsilon_{\bmB\bmD} + y_{\bmB\bmD}
\epsilon_{\bmA\bmC}, \qquad z_{\bmA\bmC} \epsilon_{\bmB\bmD} + z_{\bmB\bmD} \epsilon_{\bmA\bmC},
&\\
& h_{\bmA\bmB\bmC\bmD}\equiv -\epsilon_{\bmA(\bmC} \epsilon_{\bmD)\bmB}. &
\end{eqnarray*}
A number of identities for the above objects can be found in
\cite{FriKan00}. It is noticed that only
$\epsilon^2_{\bmA\bmB\bmC\bmD}$, $x_{\bmA\bmC}
\epsilon_{\bmB\bmD} + x_{\bmB\bmD} \epsilon_{\bmA\bmC}$ and
$h_{\bmA\bmB\bmC\bmD}$ have spin weight $0$.



\end{document}